\documentclass{mn2e}

\usepackage{astrojournals}
\usepackage{amsmath}
\usepackage{graphicx}
\usepackage{times}
\usepackage{url}
\usepackage{amssymb}
\usepackage{natbib}
\usepackage[]{hyperref}

\newcommand{\vmax}{V_{\rm max}}
\newcommand{\kms}{\mathrm{km~s^{-1}}}
\newcommand{\cmg}{\mathrm{cm^2~g^{-1}}}

\newcommand{\mstar}{M_\star}
\newcommand{\msun}{M_\odot}

\newcommand{\mvir}{M_\mathrm{v}}
\newcommand{\rvir}{R_\mathrm{v}}
\newcommand{\vhalf}{V_\mathrm{1/2}}
\newcommand{\rhalf}{r_\mathrm{1/2}}

\newcommand{\sigmam}{\sigma/m}

\title{Core Formation in Dwarf Halos with Self Interacting Dark Matter: No Fine-Tuning Necessary}

\author[Elbert et al.]{Oliver D. Elbert$^1$\thanks{$\!$oelbert@uci.edu},
	James S. Bullock$^1$,  
	Shea Garrison-Kimmel$^1$,
	Miguel Rocha$^2$, \and
	Jose O\~{n}orbe$^{1,3}$, 		
	Annika H. G. Peter$^4$ \\
	\noindent$\!$ $^1$Center for Cosmology, Department of Physics and Astronomy,
	  University of California, Irvine, CA 92697, USA \\
	    \noindent$\!\!$ $^2$SciTech Analytics, Inc., Santa Cruz, CA, 95062, USA\\
	  \noindent$\!\!$ $^3$Max Planck Institut fuer Astronomie, Koenigstuhl 17, 69117 Heidelberg, Germany \\
	  \noindent$\!\!$ $^4$CCAPP, Department of Physics, and Department of Astronomy, The Ohio State University, Columbus, OH, 43210, USA \\}
\begin{document}

\pagerange{\pageref{firstpage}--\pageref{lastpage}} 
\pubyear{2014}

\maketitle
\date{\today}

\label{firstpage}

\begin{abstract}
We investigate the effect of self-interacting dark matter (SIDM) on the density 
profiles of  $\vmax \simeq 40 ~\kms$ isolated dwarf dark matter halos -- the scale of relevance for the too big to fail problem (TBTF)  -- using very high-resolution cosmological zoom simulations.  Each halo has millions of particles 
within its virial radius. We find that SIDM models with cross sections per unit 
mass spanning the range $\sigma/m = 0.5 - 50~\cmg$  alleviate TBTF and produce 
constant density cores of size $300-1000$ pc, comparable to the half-light radii of
$\mstar \sim 10^{5-7} \msun$ dwarfs.  The largest, lowest density cores develop for
cross sections in the middle of this range, $\sigma/m\sim 5-10~\cmg$.  Our largest 
SIDM cross section run ($\sigma/m = 50~\cmg$) develops a slightly denser core owing 
to mild core-collapse behavior, but it remains less dense than the CDM case and retains 
a constant density core profile.  Our work suggests that SIDM cross sections as large 
or larger than $50~\cmg$ remain viable on velocity scales of dwarf galaxies 
($v_{\rm rms} \sim 40~\kms$).  The range of SIDM cross sections that alleviate TBTF 
and the cusp/core problem spans at least two orders of magnitude and therefore need 
not be particularly fine-tuned.
\end{abstract}


\section{Introduction}
Cosmological studies of the large-scale universe have provided tremendous 
evidence in favor of a Universe dominated by dark matter (DM) and dark energy
\citep[e.g.][]{Komatsu11,WMAP9,PlanckCos}, but thus far very little is known 
about the underlying nature of  the DM particle other than that it is long-lived 
and that it interacts weakly with the standard model.  For thermal particles, the 
dark matter needs to be fairly massive (or ``cold" -- non-relativistic at decoupling) 
in order to produce a power spectrum consistent with large-scale structure 
\citep{Reid10}.  Weakly interacting massive particles (WIMPs), for example,
provide a compelling and well-motivated class of CDM candidates 
\citep{SteigmanTurner85,Griest88,Jungman96}.  On the scales of concern for 
galaxy formation, WIMPs behave as collisionless particles and are one of the 
prime motivations for what has become the standard paradigm for structure 
formation: collisionless Cold Dark Matter (CDM).

There are, however, disagreements between predictions from CDM-only simulations 
and observations on small scales.  For example, some galaxies appear to have 
flat central density profiles \citep[e.g.][but also see \citealp{Strigari14}]{Flores1994,KuziodeNaray08,Amorisco2014,Oh08,Walker11} with 
core-like log-slopes ($\alpha \sim 0$) rather than the predicted cusps 
($\alpha \sim 1$) seen in CDM simulations \citep[][]{Dubinski1991,NFW}.  This 
issue is known as the cusp/core problem.  A qualitatively similar anomaly was 
pointed out by \citet{MBK11,MBK12}, who showed that observations of the dwarf 
spheroidal (dSph) satellites of the Milky Way (MW) imply central masses well 
below those of the $\vmax = 40~\kms$  halos that appear to be common in 
ultra-high resolution simulations of MW-size hosts 
\citep[e.g.][]{Diemand2008,Aquarius}.  This issue is known as the ``too big to 
fail" problem (TBTF).  \citet{Tollerud14} and \citet{GarrisonKimmelTBTF} further 
showed that the Andromeda (M31) dSph satellites and the field galaxies near the 
MW and M31, respectively, suffer from the same problem.  Studies of larger samples 
of dwarf galaxies in the field also indicate a similar density problem at a 
comparable velocity scale $\vmax \simeq 40 ~\kms$ 
\citep{Ferrero2012,Klypin2014,Papastergis14}.

One natural solution to the TBTF problem and other central-density issues
is to posit that galaxy halos have lower core densities than predicted in 
CDM-only simulations. For example, baryonic processes may act to reduce and 
flatten the central densities of small galaxies in CDM
\citep[e.g.][]{Pontzen12,Governato12,DiCintio14,Navarro96}. In contrast, 
\citet{Penarrubia12} and \citet{SGK13} have argued that the 
$\mstar \sim 10^6 \msun$ galaxies of interest for TBTF have not had enough 
supernovae to alter densities sufficiently 
\citep[though see][]{Amorisco2014,Gritschneder2013}.  Others have suggested 
environmental effects may increase the efficacy of these internal processes 
by similarly reducing the central masses of subhalos 
\citep[e.g.][]{Zolotov2012,DelPopolo2012,Arraki14,Brooks2014,DelPopolo2014}.
However, these external processes are weak or non-existent in the field, 
suggesting that non-baryonic solutions may be necessary to solve the 
discrepancy observed outside the MW and M31 
\citep{GarrisonKimmelTBTF,Papastergis14}.  Similarly, while large TBTF subhalos 
become increasingly infrequent as the mass of the MW and M31 host halos decrease 
\citep{Purcell12,Wang2012,Rodriguez-Puebla13,Cautun2014},
solutions of this kind appear less likely in the face of evidence that the TBTF 
problem is persistent in the field \citep{Ferrero2012,Klypin2014,GarrisonKimmelTBTF}.

Alternatively, the central density issues may be telling us something about
cosmology.  For example, if the primordial 
power spectrum is non-standard \citep{ZB02,PolisenskynRicotti,GKBicep} or 
the dark matter is warm rather than cold \citep{Anderhalden13, Lovell14,Horiuchi14} 
then the central densities of dark matter halos would decrease.  However, 
neither of these possibilities produce constant-density cores on observable 
scales; they simply lower the normalization while retaining cuspy slopes.

The focus of this paper is to explore an alternative possibility: that
the CDM particles are strongly self-interacting.\footnote{SIDM models 
with primordial power spectra that deviate from CDM on small scales have 
been explored by \cite{Buckley2014} but we focus on CDM-type power spectra 
here.}  First discussed in an astrophysical context by \citet{Spergel00}, 
Self-Interacting Dark Matter (SIDM) with energy-exchange cross sections 
per unit mass of order $\sigma/m \sim 1~\cmg$ can naturally produce 
constant-density cores in the hearts of dark matter halos on the scales of 
relevance for galaxy rotation curves and the TBTF problem 
\citep[][]{Burkert2000,Dave2001,Vogelsberger12,Rocha13}.  
Constant density isothermal cores in SIDM halos come about because
kinetic energy can be transmitted from the hot outer halo
inward \citep[see, e.g., the discussion in][]{Rocha13}. This effect can only 
occur if $\sigmam$ is large enough for there to be a relatively high
probability of scattering over a time $t_{\rm age}$ comparable to the age of 
the halo:  $\Gamma \times t_{\rm age} \sim 1$, where 
$\Gamma \simeq \rho_{\rm dm} (\sigma/m) v_{\mathrm{rms}}$ is the interaction 
rate, $\rho_{\rm dm}$ is the local dark matter density, and $v_{\mathrm{rms}}$ 
is the rms speed of dark-matter particles.  If the cross section is too large, 
however, then the high probability of interaction can potentially lead to a 
negative heat flux, where energy is transmitted from the inside out 
\citep{Kochanek2000}, leading to a ``core collapse" phenomena, not unlike 
core collapse in globular clusters,  where the central halo density 
increases to the point that it exacerbates the over-density problem on 
small scales.  The specific range of cross sections that are a) small 
enough to be observationally allowed, b) large enough to alleviate 
the relevant small-scale problems, and c) small enough to avoid catastrophic 
core collapse, are all topics of this exploration.

In most particle-physics based models for SIDM, the scattering interactions have
a velocity dependence \citep[e.g.,][]{Loeb11,TulinYuZurek,Kaplinghat14,Boddy14b,Boddy14a}.
Astrophysical constraints, on the other hand, tend to rely on specific classes of 
objects that have a characteristic $v_{\mathrm{rms}}$ scale, meaning that they 
constrain $\sigma(v)/m$ at a specific value of $v \simeq v_{\mathrm{rms}}$.  
The best published limits ($\sigma/m \lesssim 1~\cmg$) come from galaxy clusters
with characteristic rms velocities $v_{\mathrm{rms}} \simeq 1000 \, \kms$
\citep{Yoshida00b,Gnedin01,Randall08,Rocha13,Peter13}.  At the same time, the 
most compelling astrophysical {\em motivations} for exploring SIDM in the first 
place occur on the scales of dwarf galaxies where 
$v_{\mathrm{rms}} \simeq 10-100 \, \kms$. These differences in velocity scale 
are significant.  For example, the SIDM cross section could vary as
$\sigma/m \propto v^{-4}$ if the interaction were a dark version of Rutherford 
scattering, with a massless force carrier \citep{Tulin13}.  Thus it is not outlandish
to consider the possibility that DM self-interactions on the scale of dwarf galaxies 
are {\em four  orders of magnitude larger} than they are on the scale of galaxy clusters.  
Clearly we have significant need to derive constraints on as many velocity scales 
as possible.

The potential for significant velocity scaling in the SIDM energy-exchange 
cross section has motivated \citet{Vogelsberger12} and \citet{Zavala13} to 
explicitly run zoom simulations of Milky Way-size hosts using SIDM with 
velocity-dependent $\sigmam$ values tuned to evade bounds on cluster scales 
and to have large values ($\sim 10~\cmg$) on the scale of problematic TBTF 
halos ($v_{\mathrm{rms}} \simeq \vmax \simeq 40 \, \kms$).  They show that 
the TBTF problem is resolved for $\sigma/m \simeq 1-10~\cmg$ on the velocity 
scale of dwarfs.  We note that Figure~8 of \citet{Vogelsberger12} makes clear 
that a run with $\sigma/m = 10~\cmg$ provides a particularly good match 
to the spread in dwarf satellite central densities seen around the Milky Way.

The goal of this paper is to explore more fully a range of cross sections at 
the dwarf scales of interest, focusing specifically on isolated halos rather 
than subhalos in order to achieve very high resolution.  We run a set of 
ultra-high resolution cosmological simulations of isolated dwarf halos with 
$v_{\mathrm{rms}} \simeq 40 \, \kms$ using SIDM cross sections 
$\sigma/m = 0.1$ -- $50 ~\cmg$ in addition to collisionless CDM.  The aim is 
to quantify the range of cross sections that can alleviate the TBTF problem 
(in the field) and are expected to produce observable cores in small dwarf 
galaxies.  We also investigate whether a certain range of cross sections can 
be ruled out because they would result in catastrophic core collapse.

This work is organized as followed.  In \S\ref{sec:simulations}, we describe 
the simulations and analysis pipeline.  We present our results in 
\S\ref{sec:results}, focusing first on the impact of varying $\sigma/m$ on the 
density profiles in \S\ref{ssec:profiles} and then on the implications for TBTF 
in \S\ref{ssec:TBTF}.  We summarize our results and conclude in 
\S\ref{sec:conclusions}.

\begin{table}
\centering
\begin{tabular}{lccccc}
Name & $\mvir$ & $\rvir$ & $\vmax$ & $N_\mathrm{p}(\rvir)$ & $\sigma/m$ \\ 
		  &	  ($10^{10} \msun$) & (kpc) & ($\kms$) & 	($10^6$) & ($\cmg$)	\\ \hline \hline
Pippin & $0.9$  & $55$ & 37 & $4.1$ & $0, 0.1, 0.5,  $ \\
	  &		&	    &	    &		&  $ 5,10, 50 $  \\ 
\hline
Merry & $1.2$  & $59$ & 38 & $5.4$ & $0, 0.5, 1, 10$ \\ \hline
\end{tabular}
\caption{Summary of simulated halos. The first four columns list identifying 
names and virial-scale properties (virial mass, virial radius, and maximum 
circular velocity).  The fifth column gives number of particles 
within the virial radius for the high-resolution runs and the last column 
summarizes the cross sections each halo was simulated with. The virial-scale 
properties of the halos listed are for the CDM cases ($\sigma/m = 0$) but each 
of these values remains unchanged (within $\sim 5\%$) for all SIDM runs.  $\mvir$ and $\rvir$ are calculated using the \citet{Bryan98} definition of $\rho_{\mathrm{V}}$.}
\label{sims.tab}
\end{table}

\section{Simulations}
\label{sec:simulations}

\begin{figure*}
\centering
\includegraphics[width = \columnwidth,angle=-90]{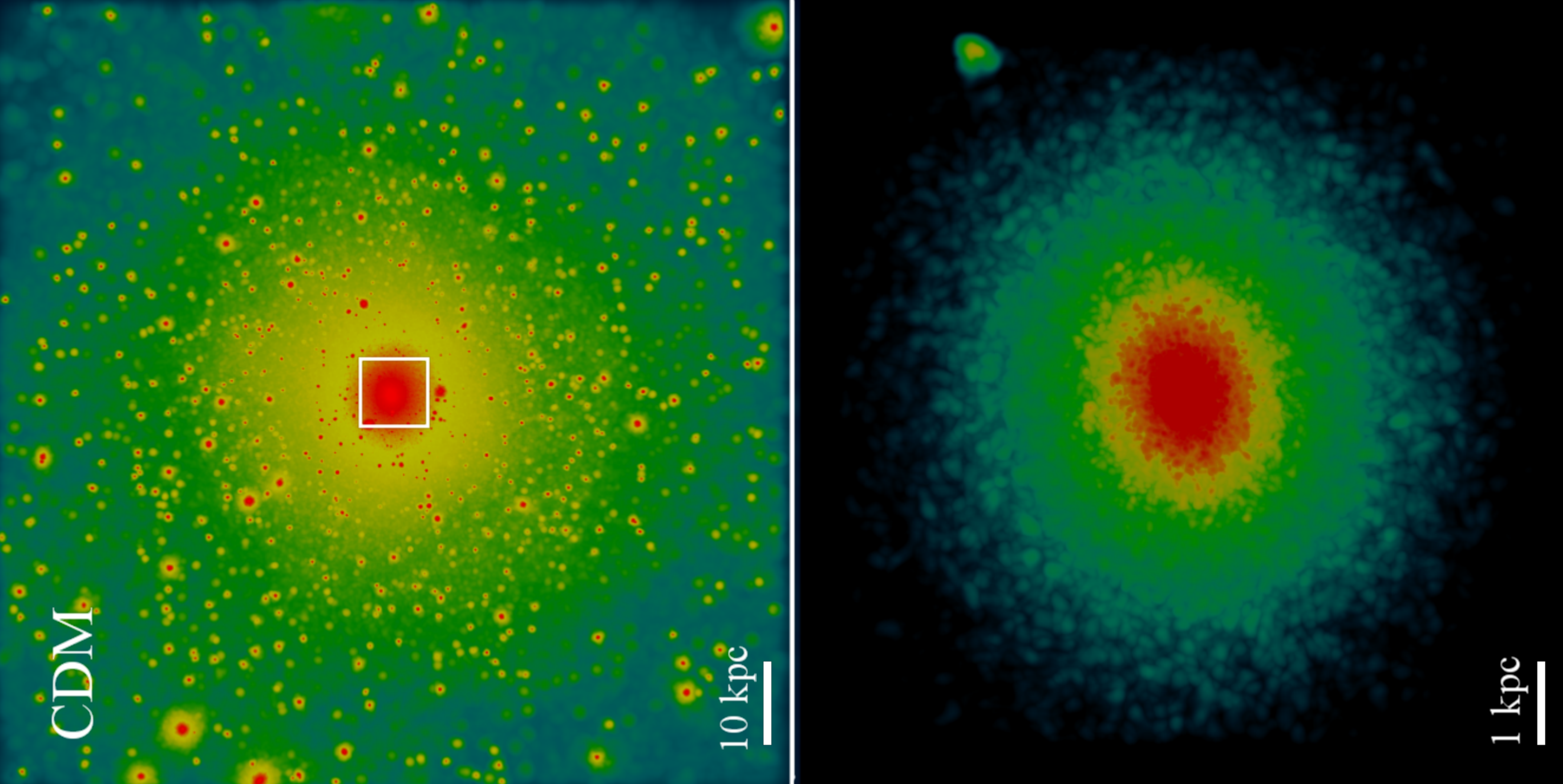} 
\includegraphics[width =  \columnwidth,angle=-90]{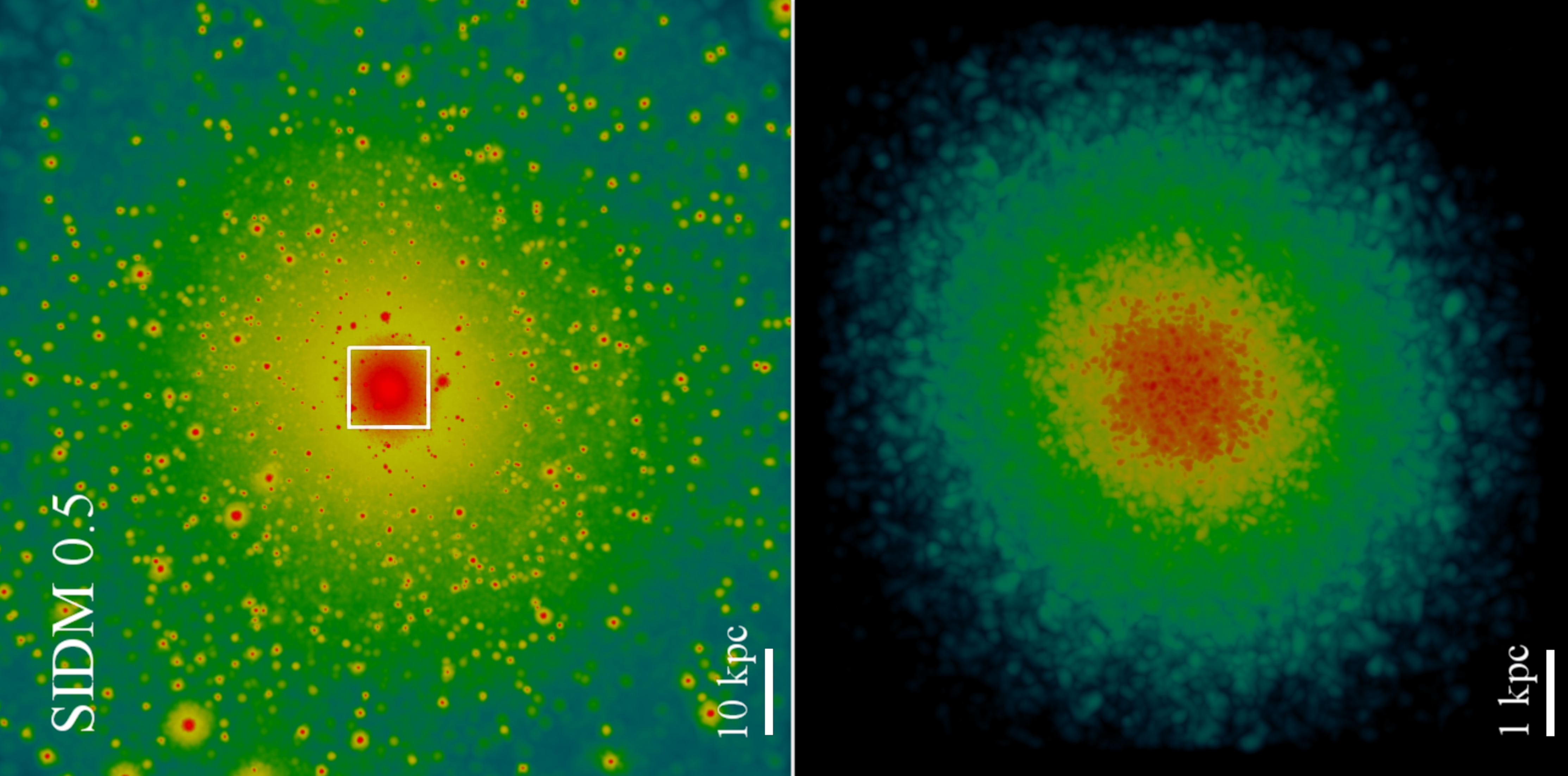}
\includegraphics[width =  \columnwidth,angle=-90]{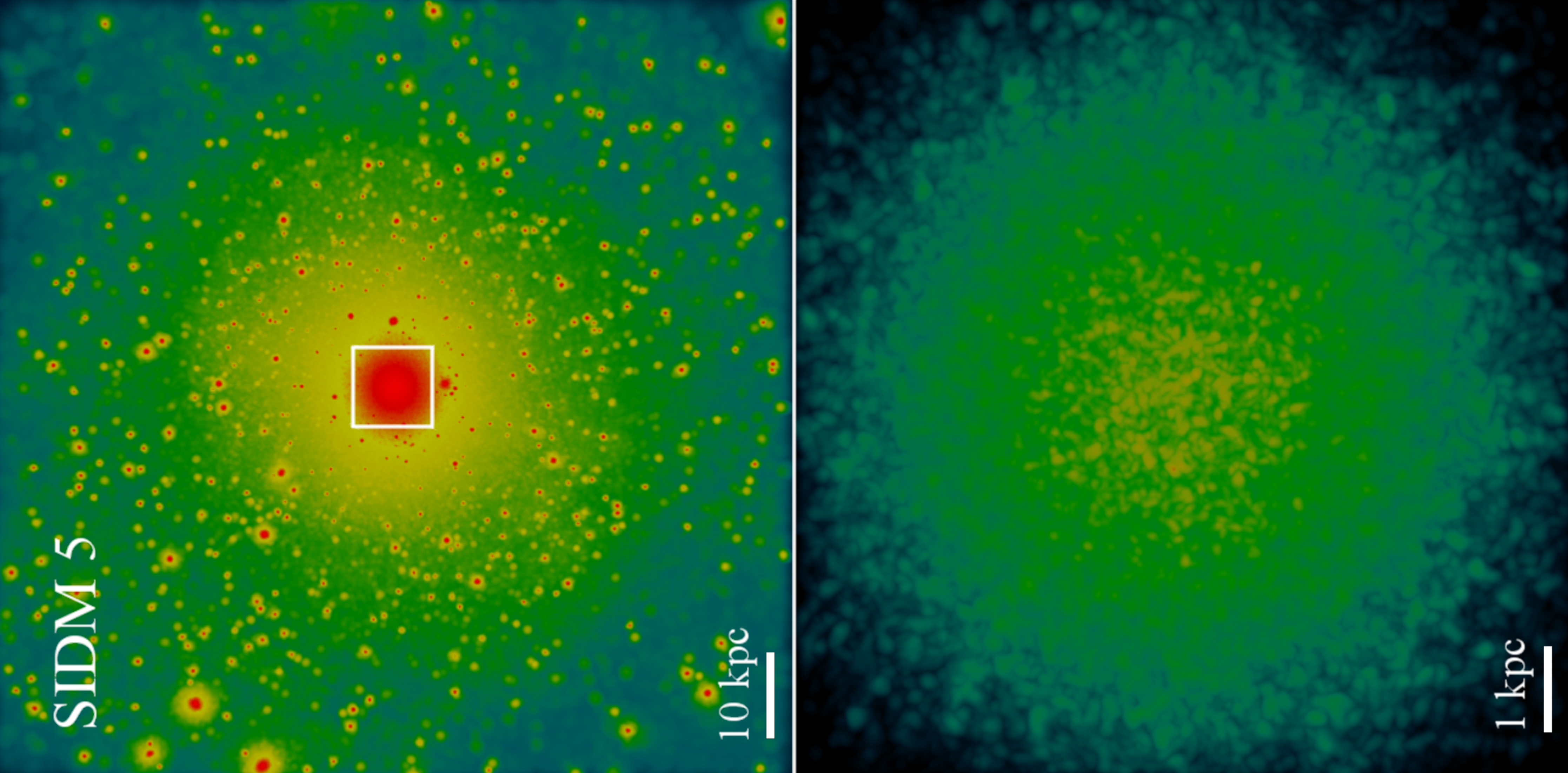}
\includegraphics[width =  \columnwidth,angle=-90]{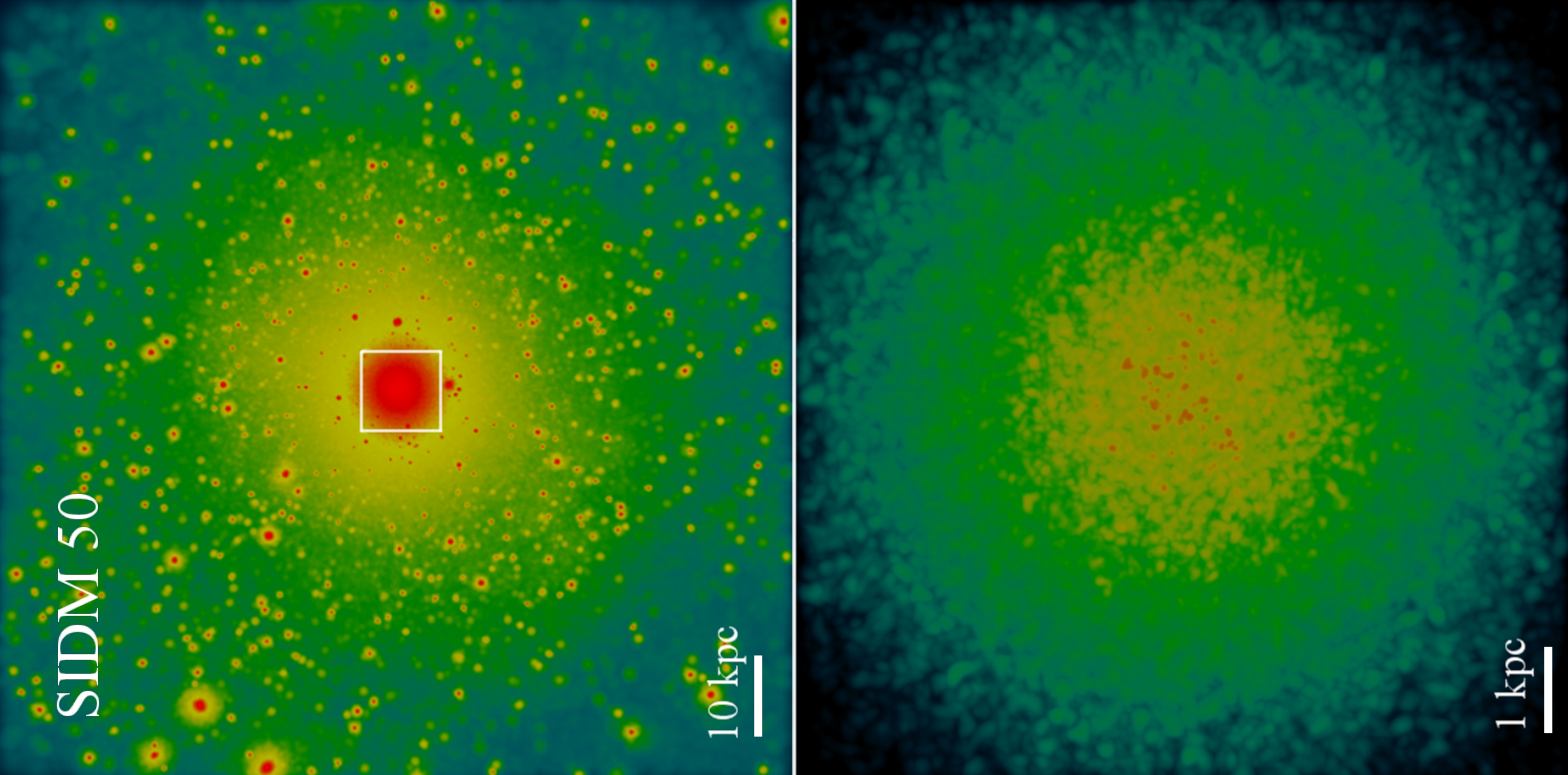}
\caption{Dark matter density of Pippin in CDM (left) and SIDM with $\sigma/m$ increasing from left to right: 0.5, 5, and 50 $\cmg$. Boxes on the top span 100~kpc ($\rvir = 55 \mathrm{kpc}$)  and the
bottom panel zooms in to span a central 10~kpc box (with modified color bar).  Notice that the 
 global properties of the halos on the scale of the virial radius, including the number and locations of subhalos, are
 nearly identical across all runs.  The only difference is that the inner core regions become less dense and somewhat puffed out  in the SIDM cases.  Note that the $50~\cmg$ simulation is somewhat denser in the inner core than the $5~\cmg$ case; it is undergoing mild core collapse.}
\label{fig:vizes}
\end{figure*}

\begin{figure}
\centering
\includegraphics[width = \columnwidth]{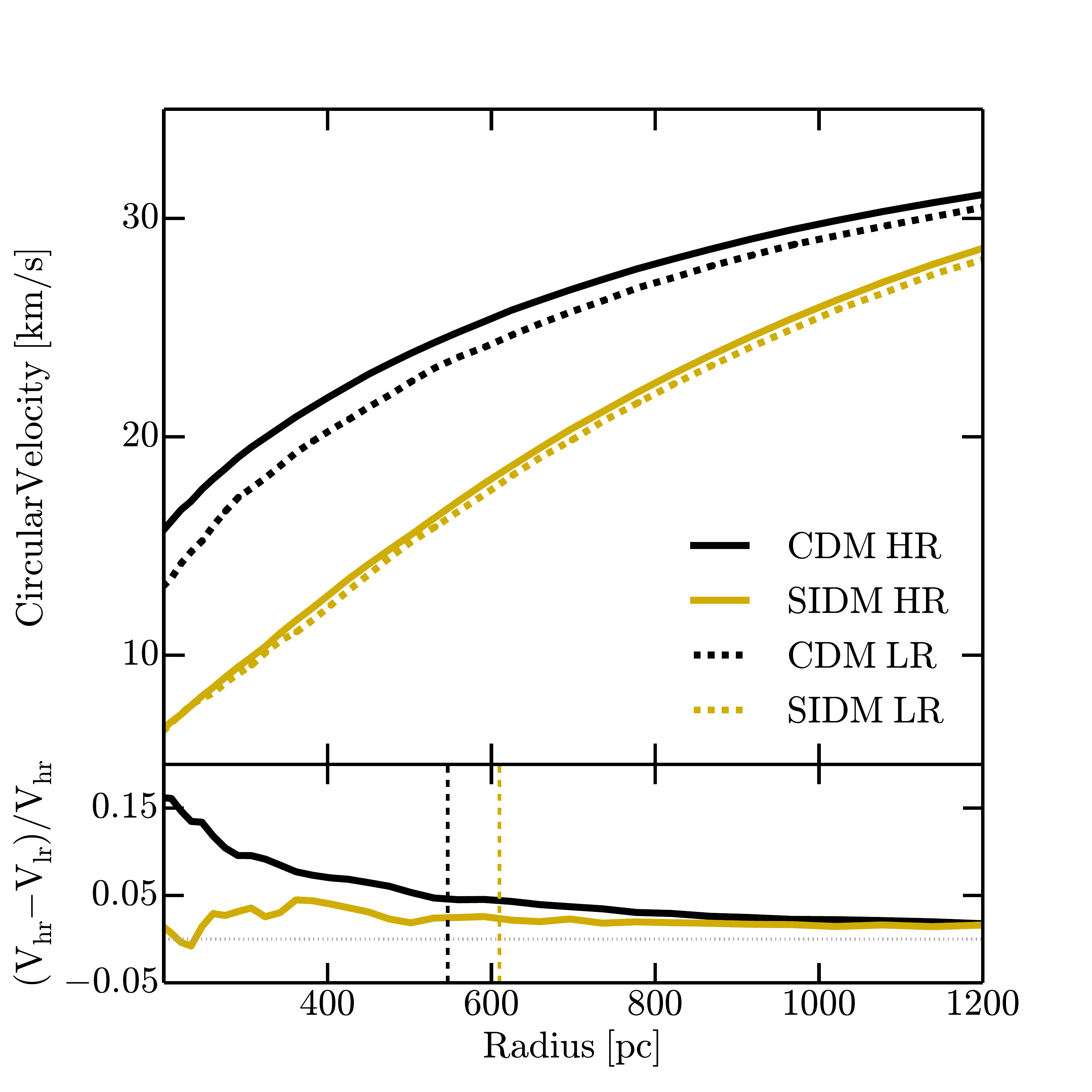}
\caption{\textit{Top:} Circular velocity profiles of Pippin at our fiducial resolution (solid; $m_\mathrm{p} = 1.5\times10^3~\msun$, $\epsilon = 28$~pc)  and at low resolution (dotted; $m_\mathrm{p} = 1.2\times10^4~\msun$,  $\epsilon = 84$~pc)
for both CDM (black) and SIDM (1 $\cmg$, yellow).  \textit{Bottom:} The relative difference between high and low resolution circular velocity profiles.  The dashed vertical lines indicate the \citet{Power03} radius for the low resolution halos. Note that the SIDM halos are much better converged than the CDM simulations.}
\label{vcircconvergence.fig}
\end{figure}

As we demonstrate explicitly below, in order to resolve the relevant 
$\sim 300 - 1000$~pc scales of classical dwarf spheroidal galaxies, very high 
force and mass resolution are required.  We achieve this resolution in cosmological
simulations using the zoom technique  \citep{Katz1993,Onorbe14}.  Our SIDM 
implementation follows that described in \citet{Rocha13}, using a modified version of 
\texttt{GADGET-2} \citep{Springel05}. Halos were identified with the six-dimensional 
phase-space halo finder \texttt{ROCKSTAR} \citep{Behroozi13}.  

We chose two halos for our primary simulations using parent cosmological volumes of 7 Mpc on a side.  Initial conditions were generated with \texttt{MUSIC} \citep{Hahn11} 
at $z = 125$ using cosmological parameters derived from the \textit{Wilkinson 
Microwave Anisotropy Probe}-7 year data \citep{Komatsu11}:  $h = 0.71$, 
$\Omega_m = 0.266$, $\Omega_{\Lambda} = 0.734$, $n_s = 0.963$, and 
$\sigma_8 = 0.801$.  Their global properties are given in Table~\ref{sims.tab}.  
We refer to the slightly smaller of the two dwarfs ($\vmax = 37~\kms$) as 
Pippin and the larger ($\vmax = 38~\kms$) as Merry.  Our high resolution runs, 
which we analyze throughout, have particle mass $m_\mathrm{p} = 1.5\times10^3~\msun$ 
and a Plummer equivalent force softening $\epsilon = 28$~pc.  We have also checked 
that various basic parameters of our target halos (spins, concentrations and formation 
times) are within one standard deviation of what is expected for dwarf halos 
based on a larger simulation box of 35~Mpc on a side 
\citep[described in][]{Onorbe14}.


In addition to $\sigma/m = 0$ (collisionless CDM) runs, we simulate both 
halos with $\sigma/m = 0.5, 1, 10~\cmg$.  Additionally we have simulated 
Pippin with $\sigma/m = 0.1, 5, 50~\cmg$.  In all SIDM simulations, the 
dark matter self-interactions were calculated using an SIDM smoothing length 
equal to $0.25\epsilon$, as described in \citet{Rocha13}. 

Figure~\ref{fig:vizes} shows visualizations of Pippin at high resolution, 
colored by the local dark matter density, with collisionless CDM on the far 
left and SIDM runs of increasing cross section to the right.  The upper panels 
visualize a box 100~kpc across ($\sim2\rvir$) and the lower panels zoom in 
on the central 10~kpc of the halos, using a color bar that has been rescaled to 
emphasize the highest densities. As these visualizations emphasize, bulk halo 
properties on the scale of $R_{\rm v}$ are virtually identical in CDM and SIDM;  
even the locations of subhalos remain unchanged.  The fact that substructure 
remains very similar in both SIDM and CDM is consistent with the findings of 
\citet{Vogelsberger13} and \cite{Rocha13}; here, however, we examine mass 
scales well below those resolved in any previous SIDM study, resolving 
substructure as small as $V_{\rm max} = 1~\kms$.  The main differences are apparent 
in the core regions (lower panels), where the SIDM runs are systematically less 
dense than CDM.  Note that the $50~\cmg$ run is actually denser in its core than 
the $5~\cmg$ run.  As discussed below, this is a result of core collapse.

\subsection{Resolution Tests}
\label{ssec:restest}

We have designed our high-resolution simulations explicitly to recover the 
density structure at the $\sim 300$ pc half-light radius scale of low-mass 
dwarfs based on the work of \citet{Power03} for CDM simulations.  \citet{Power03} 
showed that the differential density profiles of CDM halos should be converged 
only outside of a specific radius where the gravitational two-body relaxation 
time approximates the Hubble time.  While this work is perfectly well-designed 
for CDM runs, the issue of convergence in SIDM is less well explored.  In order 
to remedy this concern, we have simulated Pippin in CDM and SIDM 
(1~$\cmg$) at lower resolution, with eight times worse mass resolution 
($m_\mathrm{p} = 1.2\times10^4~\msun$) and with greater force softening 
($\epsilon = 84$~pc) than our high resolution runs.  As expected, we confirm that 
the differential density profile of the CDM halo is convergent down to the 
classic \citet{Power03} radius of the low resolution runs ($168$ pc for Pippin, 
$160$ pc for Merry); reassuringly, the SIDM run is even more stable.  We find 
convergence in the density profile down to below half the Power radius 
\citep[see also][who found similar robustness for SIDM halos]{Vogelsberger12}.  
This is qualitatively reasonable in the limit where physical self-interactions 
are more important than artificial two-body interactions.

While the differential density is the most natural theoretical quantity to 
consider in a convergence study, observationally the circular velocity (or 
cumulative mass) is more relevant.  Velocity curves suffer more from 
numerical convergence issues because they rely on the integrated density.  
Figure~\ref{vcircconvergence.fig} shows the circular velocity profiles of 
the low (dashed) and high resolution (solid) simulations with CDM in black 
and SIDM in yellow.  The lower panel shows the relative difference as a 
function of radius, with the Power radii of the low resolution runs marked 
for comparison as vertical dashed lines.  The low resolution rotation curve 
in CDM starts to under-predict visibly compared to the higher resolution run 
at about $1.4$ times the Power radius (some $14$ times the formal force 
softening $\epsilon$) and disagrees by more than $5\%$ at 
$\sim1.2\times \mathrm{r_{Power}}$.  The SIDM run, however, remains reasonably 
well converged throughout:  the high and low resolution simulations do not 
disagree by more than $4\%$ outside of $\sim0.6\times \mathrm{r_{Power}}$, and 
not by more than $5\%$ outside of 100 pc.

For the remainder of this work, we present high resolution density 
profiles of CDM halos down to the Power radius and to half the Power radius for SIDM profiles.  Based on the work presented in 
this section, we believe that the full regions plotted are converged. 
For rotation curves, we plot both CDM and SIDM halos to 200~pc.  This is 
the $1.2 \mathrm{r_{Power}}$ limit for the CDM curves, where we expect them to be 
correct within $\lesssim5\%$.
The SIDM rotation curves are accurate to the last plotted point.

\section{Results}
\label{sec:results}

\subsection{Halo Profiles}
\label{ssec:profiles}

\begin{figure*}
\centering
\includegraphics[width = \columnwidth]{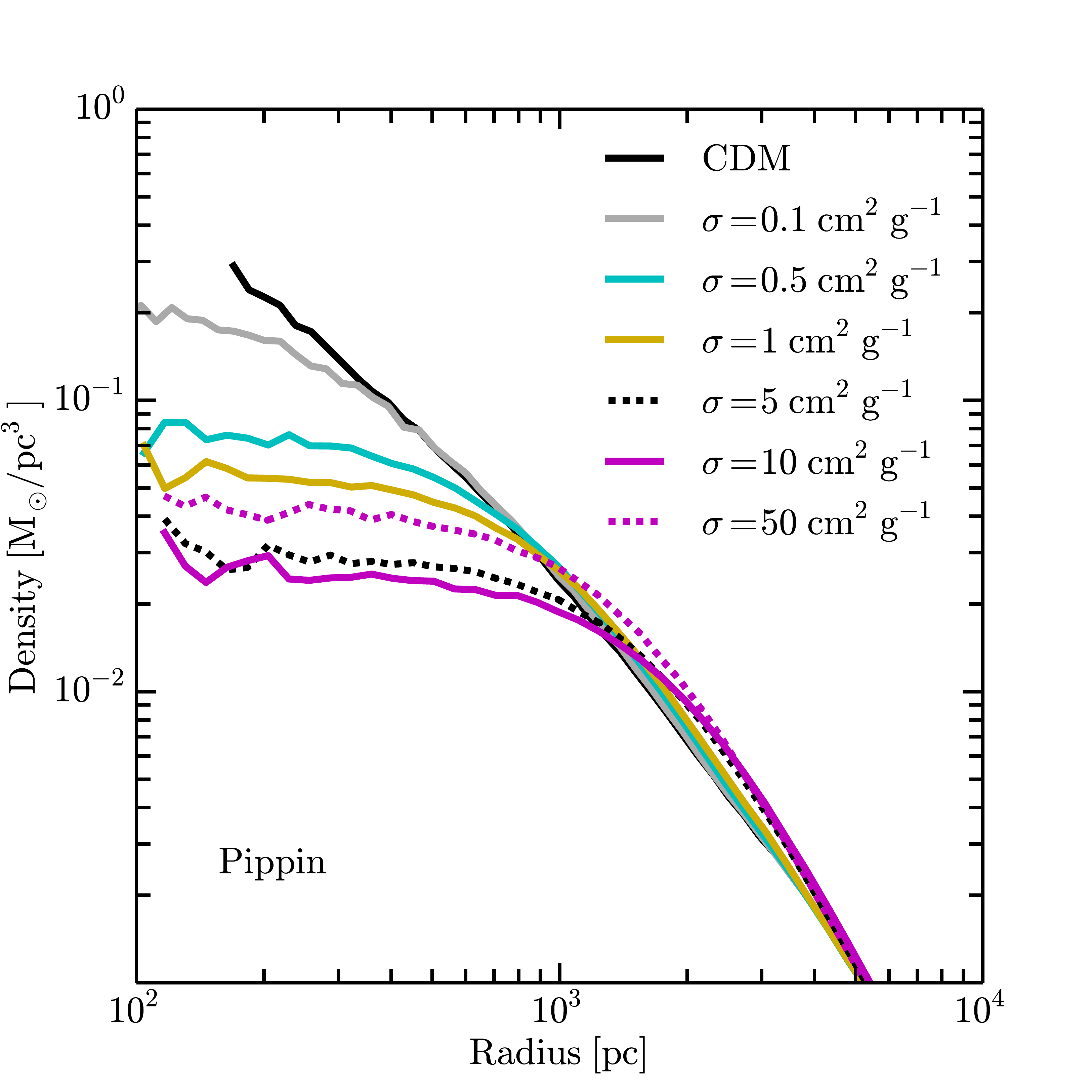} 
\includegraphics[width = \columnwidth]{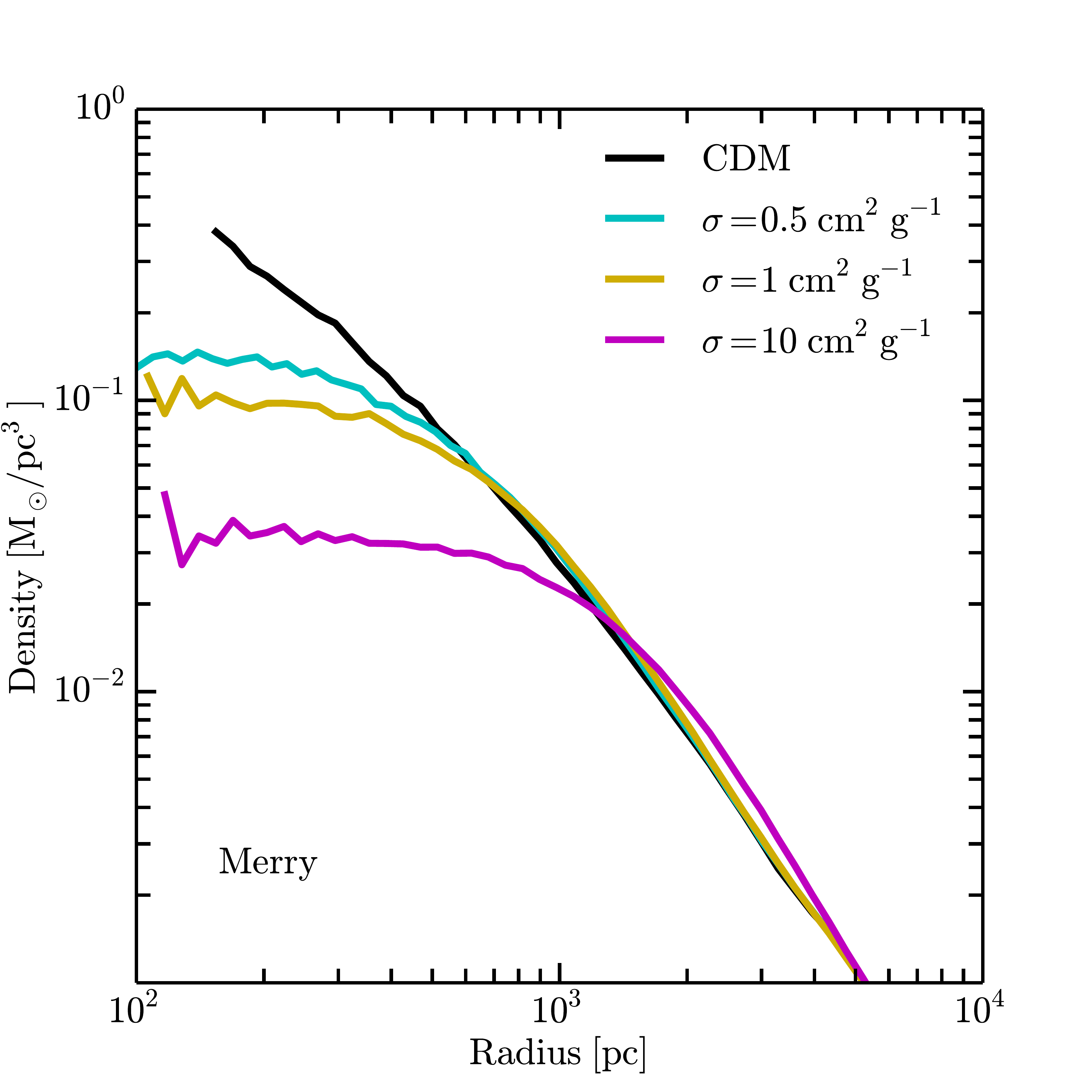} 
\caption{Density profiles of Pippin (left) and Merry (right) in collisionless CDM
and in SIDM (see legend) at $z=0$.  All SIDM runs with $\sigma/m \geq 0.5~\cmg$ 
produce central density profiles with well-resolved cores within $\sim 500$ pc.  
Core densities are the lowest (and core sizes the largest) for cross sections in 
the range $\sigma/m = 5-10~\cmg$.  The $50~\cmg$ run of Pippin has undergone a 
mild core collapse, with a resultant central density intermediate between
the $10~\cmg$ run and $1~\cmg$ run.  For velocity dispersion profiles of these 
halos, see Appendix~\ref{appendixA}.  NFW fits to the CDM profiles of each halo yield scale radii of $\sim 2.7$ kpc.}
\label{fig:profiles}
\end{figure*}

Figure~\ref{fig:profiles} presents the density profiles of our resultant halos.  
While the $\sigma/m = 0.1~\cmg$ run of Pippin produces only a modest reduction 
in core density compared to CDM, all SIDM runs with $\sigma/m \ge 0.5~\cmg$ 
result in substantial $\sim500-1000$~pc cores, with reduced central densities 
compared to CDM.  As the SIDM cross section is increased from 
$\sigma/m = 0.1 \rightarrow 5-10~\cmg$ the cores become increasingly extended
and have lower central densities.  However, the Pippin run with 
$\sigma/m = 50~\cmg$ has a denser core than the $\sigma/m = 10~\cmg$ case.  This 
is almost certainly due to core-collapse behavior  
\citep[e.g.][]{Kochanek2000,Balberg02,Colin02,Koda11}.  As illustrated and discussed in 
Appendix~\ref{appendixA}, the velocity dispersion profile of this run is noticeably 
hotter in the core than in the outer regions -- a clear indication that a negative 
heat flux is in action.  However, note that even a cross-section as large as 
$50~\cmg$ results in a significantly lower central density than the CDM case, with 
a clear constant-density core within $\sim 500$ pc.  Evidently, for this particular 
halo at least, a cross section as large as $50~\cmg$ does not produce run-away core 
collapse, but rather a mild increase in the central density compared to a run with 
ten times weaker self-interaction ($5~\cmg$).

Our results on core collapse are not significantly different than those reported 
most recently in the literature.  Older simulations of isolated halos showed core recollapse after a few dynamical times for 
an equivalent cross section larger than $\sim 5~\cmg$ \citep[scaled appropriately to the 
mass of Pippin;][]{Kochanek2000,Balberg02,Koda11}.  However, cosmological halos only see core collapse for higher cross sections, and not consistently for a fixed halo mass \citep{Yoshida00b,Colin02,Dave2001,Rocha13}. 
The primary reason for this difference is that accretion of new matter onto the halo stabilizes the heat flow within the halo \citep{Yoshida00b,Colin02}.  
\citet{Vogelsberger12} saw evidence for core collapse only for their largest 
constant cross section run ($\sigmam = 10 ~\cmg$) and, even then, only for 
one subhalo (out of the $\sim 10$ most massive they studied).  This is broadly consistent 
with the behavior we see. 

\subsection{Circular Velocities and the Too Big to Fail problem}
\label{ssec:TBTF}

According to the ELVIS simulations of the Local Group \citep[][]{ELVIS}, there 
should be $\sim 10$ isolated halos with $V_{\rm max} \gtrsim 40~\kms$ in the local 
($\sim 1.2$ Mpc) field around the MW and M31, excluding satellites of either large 
system.  Of the fourteen isolated dwarfs in this volume, only one (Tucana) is clearly 
dense enough to reside in a CDM halo larger than $40~\kms$.  The rest appear to 
reside in halos that are significantly less dense than expected for the ten most 
massive systems predicted in CDM simulations.  These missing, or overdense, 
$V_{\rm max} \simeq 40~\kms$ halos are the systems of concern for the 
TBTF problem. 

\begin{figure*}
\centering
\includegraphics[width =\columnwidth]{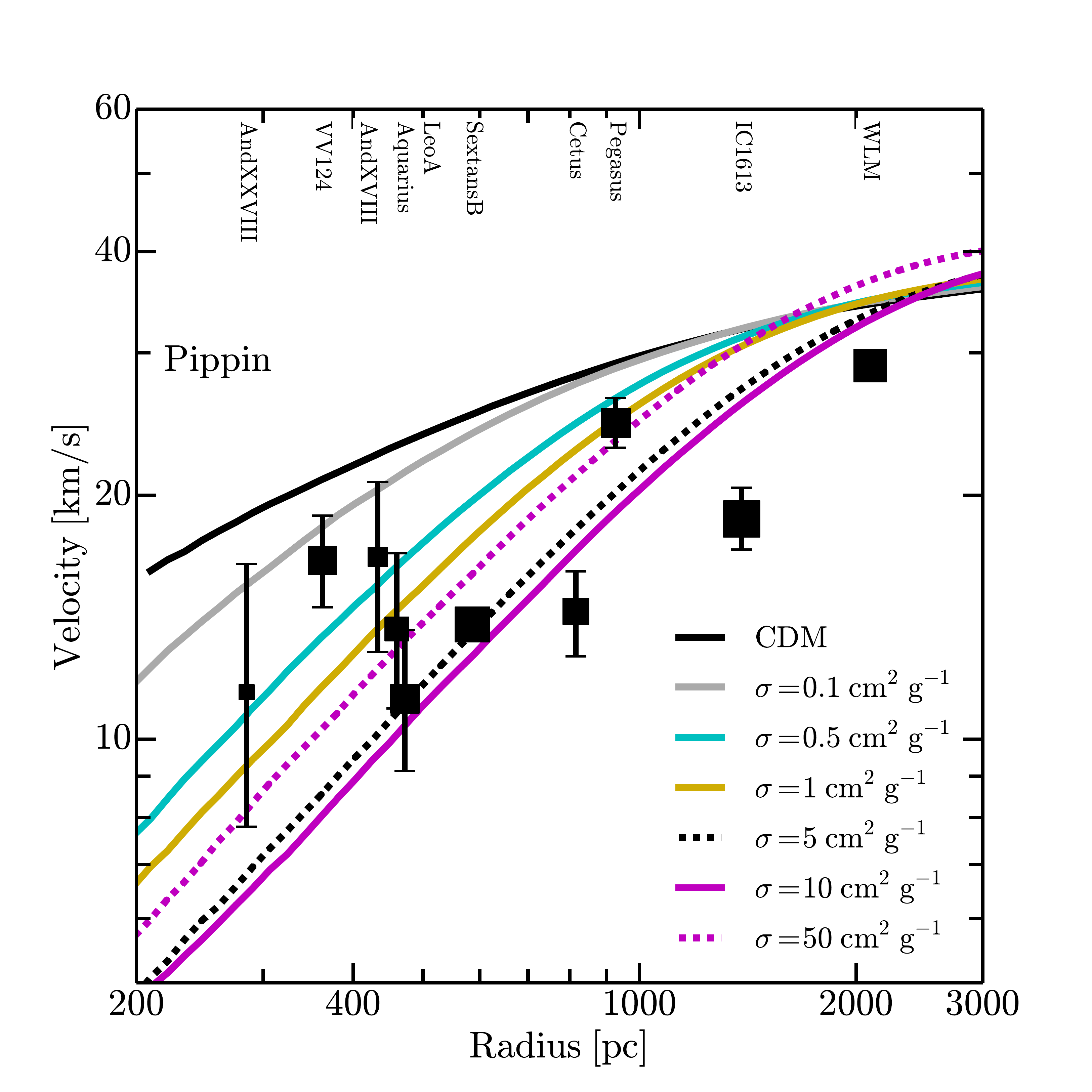}
\includegraphics[width =\columnwidth]{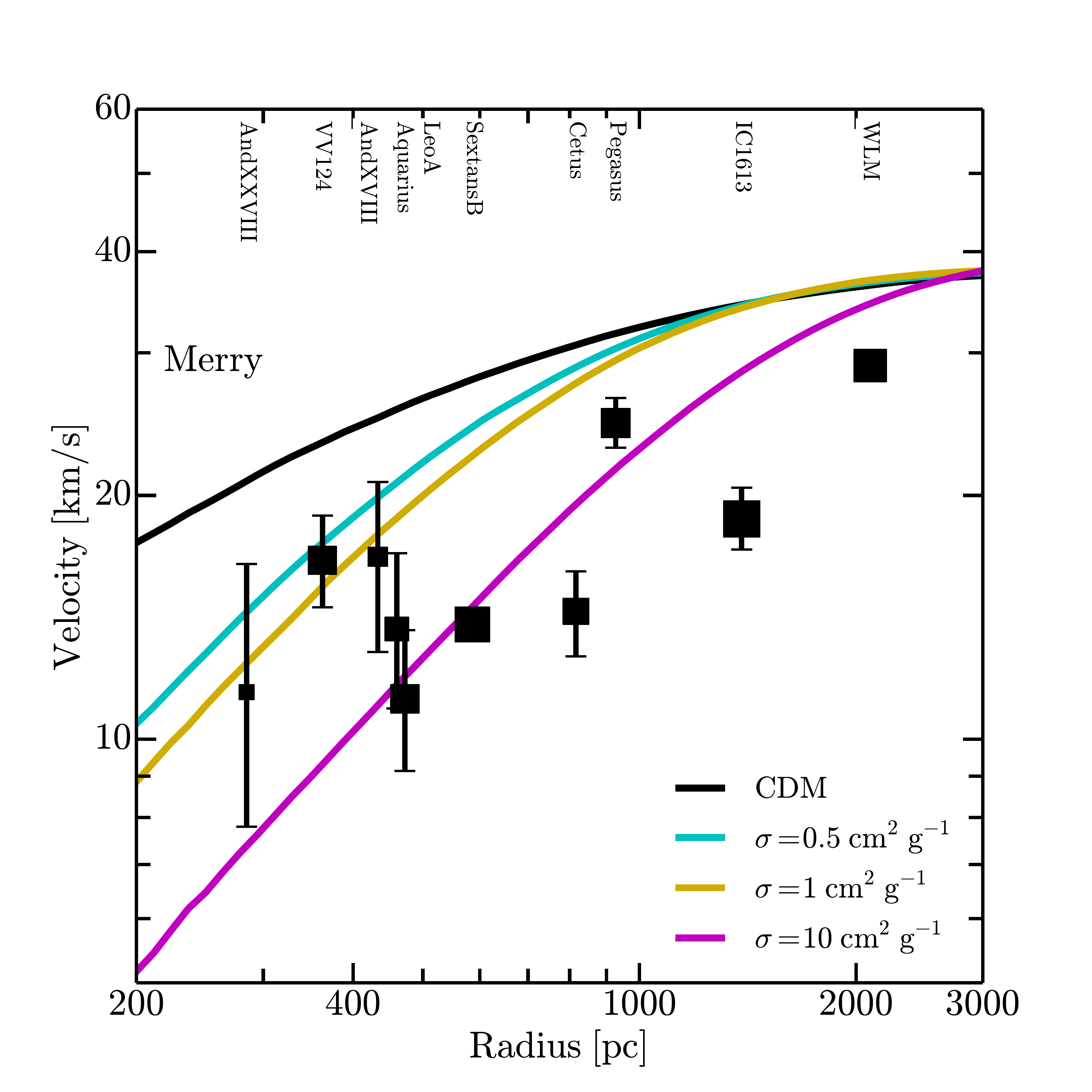}
\caption{Circular velocity profiles of Pippin (left) and 
Merry (right) simulated with CDM (solid black) and various SIDM cross-sections 
(see legend).  Halos of this mass ($V_{\rm max} \simeq 40~\kms$) should be 
fairly common within the $\sim 1$ Mpc Local Field of the Milky Way.  The data 
points indicate measurements of the circular velocity at the half-light radii 
of Local Field \citep[as defined in][]{GarrisonKimmelTBTF} dwarf galaxies (see references in text).  The sizes of the points scale with the stellar mass of 
the galaxy, with the smallest AndXXVIII at $\mstar \simeq 3 \times 10^5 \msun$ 
and the largest IC 1613 at $\mstar \simeq 2 \times 10^8 \msun$.  While these 
$V_{\rm max} \simeq 40~\kms$ halos are too dense in CDM to host any of of the 
plotted Local Field dwarfs, the same halos in SIDM with $\sigma/m \ge 0.5 ~\cmg$ 
are consistent with many of the data points.}
\label{fig:rotationcurves}
\end{figure*}

Figure~\ref{fig:rotationcurves} illustrates this problem explicitly by comparing 
the circular velocities of nearby field dwarfs at their half-light radius (data 
points) to the circular velocity profiles of our simulated halos (lines), each of 
which has $V_{\rm max} \simeq 40~\kms$ and is therefore nominally a TBTF halo.  The 
data points indicate dwarf galaxies ($\mathrm{M}_\ast < 1.7\times 10^{7}$) farther than 300~kpc 
from both the Milky Way and Andromeda that are dark matter dominated within their half-light 
radii ( $\rhalf$), with estimates for their circular velocities at $\rhalf$ ($\vhalf$).  We have 
excluded Tucana, which has an implied central density so high that it is hard to 
understand even in the context of CDM \citep[see][for a discussion]{GarrisonKimmelTBTF}.  
$\vhalf$ for the purely dispersion galaxies are calculated using 
the \citet{Wolf10} formula, where measurements for stellar velocity dispersion, 
$\sigma_\star$, are taken from \citet{Hoffman96}, \citet{Simon2007}, \citet{Epinat08}, \citet{Fraternali09}, 
\citet{Collins13}, and \citet{Kirby14}.  However, WLM and Pegasus also display
evidence of rotational support, indicating that they are poorly described by
the \citet{Wolf10} formalism.  For the former, we use the \citet{Leaman12} 
estimate of the mass within the half-light radius, obtained via a detailed
dynamical model.  The data point for Pegasus is obtained via the method
suggested by \citet{Weiner2006}, wherein $\sigma_\star^2$ is replaced with
$\sigma_\star^2+\frac{1}{2}(v\sin i)^2$ in the \citet{Wolf10} formula, where $v\sin i$ 
is the projected rotation velocity
\citep[also see \S5.2 of][]{Kirby14}.

As expected, the data points all lie below the CDM curves (black lines), 
demonstrating explicitly that both Merry and Pippin are TBTF halos.  The 
SIDM runs, however, provide a much better match, and in fact all of the SIDM 
runs with $\sigma/m \geq 0.5~\cmg$ alleviate TBTF.

\subsection{Expectations for the stellar-mass halo-mass relation}
A problem related to TBTF, but in principle distinct from it, concerns the 
relationship between the observed core densities of galaxies and their stellar 
masses.  Specifically, there does not appear to be any correlation between 
stellar mass and inner dark matter density inferred from dynamical estimates of 
dwarf galaxies in the Local Group \citep{Strigari08,MBK12,GarrisonKimmelTBTF}.  
If dark matter halos behave as expected in dissipationless $\Lambda$CDM simulations, 
then we would expect more massive galaxies to have higher dark matter densities at 
fixed radius.  This ultimately stems from the expectation, borne out at higher 
halo masses, that more massive dark matter halos tend to host more massive galaxies.  

Consider, for example, the two galaxies Pegasus ($\rhalf \simeq 1$ kpc) and Leo A 
($\rhalf \simeq 500$ pc) in Figure ~\ref{fig:rotationcurves}.  Both of these 
galaxies have about the same stellar mass $M_\star \simeq 10^7 \msun$.   According 
to the expectations of abundance matching \citep{GarrisonKimmelTBTF}, each of these 
galaxies should reside within a $V_{\rm max} \simeq 40~\kms$ halo.  Instead, their 
central densities are such that, if their dark matter structure follows the 
CDM-inspired NFW form, they need to have drastically different potential well 
depths: $V_{\rm max} \simeq 30$ and $12~\kms$ for Pegasus and Leo A, respectively 
\citep[see Figure 12 of][]{GarrisonKimmelTBTF}.  However, if we instead interpret 
their densities in the context of SIDM, the results are much more in line with 
abundance matching expectations.

Abundance matching relations remain unchanged in SIDM because halo mass functions 
in SIDM are identical to those in CDM \citep{Rocha13}.  That is, in SIDM, just 
like CDM, we would naively expect both Pegasus and Leo A to reside in halos with 
$V_{\rm max} = 40 ~\kms$, like Pippin and Merry.  In SIDM, unlike in CDM, the
predicted density profiles allow this to happen self-consistently.
In the right-hand panel of Figure~\ref{fig:rotationcurves}  we see that 
{\em both} Pegasus and Leo A could be hosted by Merry with $\sigma/m = 10~\cmg$.  
In the left-hand panel, both galaxies are consistent with the 
$\sigma/m = 50~\cmg$ line.  Given the obvious halo-to-halo scatter and small 
number of simulations we have, it is difficult to determine which cross 
section would be favored, but it is clear that for these halos SIDM predicts central densities much more in line with naive expectations for the stellar mass 
to halo mass relation at the mass scale of dwarfs.  

We extend this analysis to smaller stellar masses in the next subsection,
where we also address whether high values of $\sigmam$ are forbidden by the 
dynamics of the Local Group dwarfs.


\subsection{Are any cross sections too large to accommodate measured densities?}

We would like to be able to rule out some range of cross sections on the velocity 
scale of dwarf galaxies by requiring that dark matter densities at least as
high as those observed can be achieved.  This is in some sense the inverse of 
the standard central-density problem: for what SIDM cross sections are galaxies 
{\em too dense}?

The densest Local Field galaxies shown in Figure 4 are And~XVIII 
($\rhalf \simeq 400$ pc) and And~XXVIII ($\rhalf \simeq 300$ pc) with average 
densities just under $0.1 ~ \msun\,{\rm pc}^{-3}$.  In practice, the mass 
uncertainties on these galaxies are so large that it will be difficult to derive 
stringent constraints.  At face value, however, the $\sigma/m = 10~\cmg$ lines do 
appear to be somewhat under-dense (by a factor of $\sim 2-3$) compared to the 
central data points.   

The difficulty in this comparison is that we expect that the core densities of 
SIDM halos will {\em increase} with decreasing $V_{\rm max}$ \citep{Rocha13}. We 
must account for this possibility in any attempt to rule out a given cross section 
based on an observed galaxy density.   

In order to estimate a $V_{\rm max}$ scale that might be reasonable for these 
galaxies we can turn to abundance matching.  These dense dwarfs have 
$M_\star \simeq 3 - 8 \times 10^5 \msun$.  According to abundance matching 
estimates \citep{ELVIS}, we expect galaxies in this stellar mass 
range to reside within $V_{\rm max} \simeq 20 - 30 ~\kms$ halos.  

\begin{figure}
\centering
\includegraphics[width = \columnwidth]{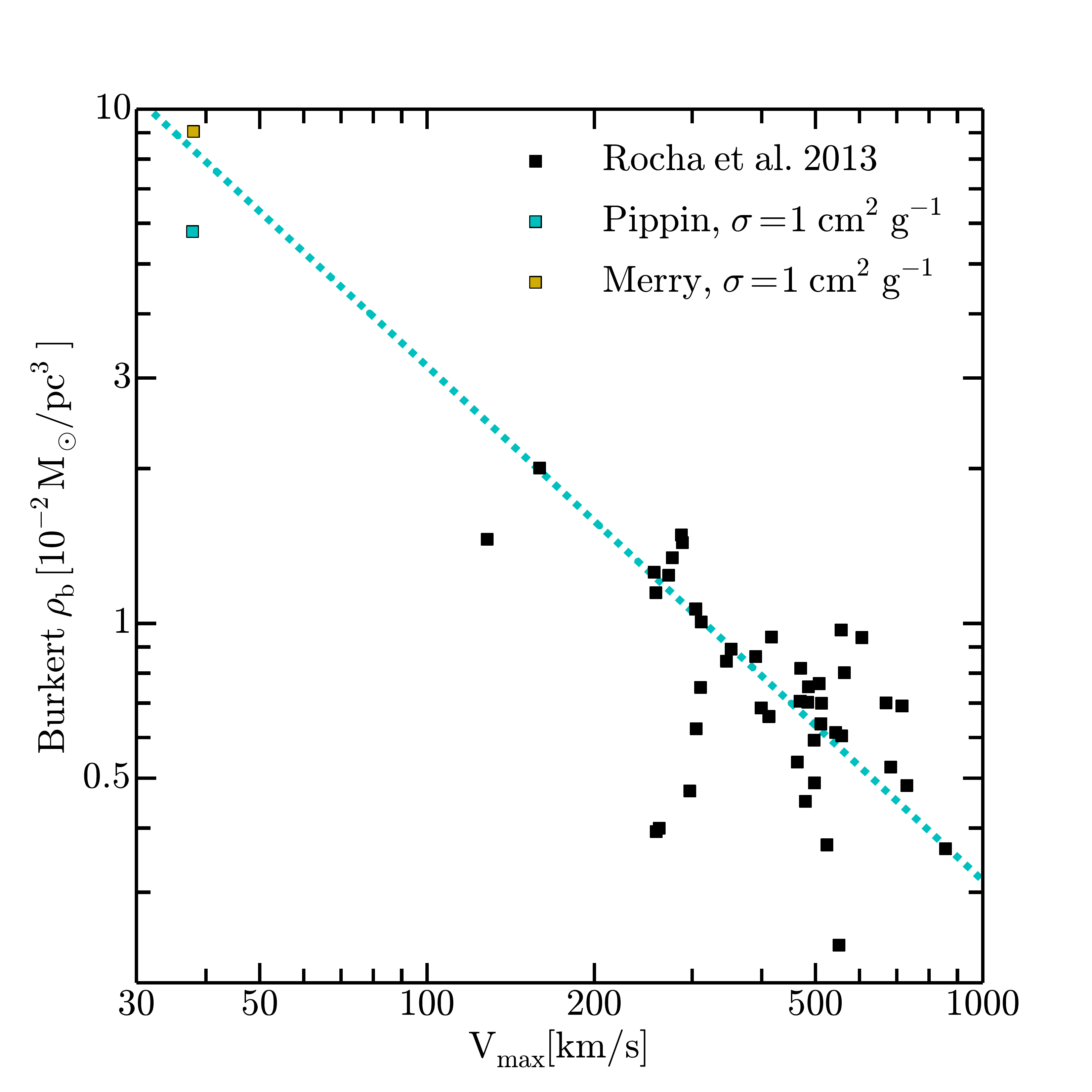}
\caption{Central halo densities of SIDM halos from fitted \citet{Burkert95} 
profiles are plotted versus the maximum circular velocities of the halos.  Black 
points are taken from \citet{Rocha13} and the colored points indicate Merry and 
Pippin with \(\sigmam=1~\cmg\).  The dotted line corresponds to a $1/V_{\rm max}$ 
scaling.  Though there is significant scatter at fixed mass, smaller halos have 
higher central densities, helping to naturally explain the ``common mass'' 
relation inferred from CDM simulations.}
\label{rhob_vmax.fig}
\end{figure}

How much denser is a  $V_{\rm max} = 20 ~\kms$ halo than a $V_{\rm max} = 40 ~\kms$ 
halo in SIDM?  Figure \ref{rhob_vmax.fig} provides some insight.   Plotted are 
fitted halo core densities (assuming a \citealt{Burkert95} profile), $\rho_{\rm b}$, 
as a function of $V_{\rm max}$ for SIDM simulations with $\sigma/m = 1 ~\cmg$.  
The black squares show halos from the simulations of \citet{Rocha13} while the 
colored squares show the two halos discussed in this paper.  The dotted line shows
a $\rho_{\rm b} \propto V_{\rm max}^{-1}$ scaling (a power law fit to all the 
plotted points yields $\rho_{\rm b} \propto V_{\rm max}^{-0.9}$).  

If the same $V_{\rm max}^{-1}$ behavior holds for core densities in 
$\sigma/m = 10 ~\cmg$ models, then the density of a  $V_{\rm max} = 20 ~\kms$ halo would be roughly comparable to the best-fit values for And XVIII and And XXVIII in 
this case.\footnote{This assumes a constant SIDM cross section.  If the cross section is instead velocity dependent this will change, most likely resulting in a shallower power law fit.}  Such a model then remains viable in the face of current constraints.    

Thus it appears difficult to rule out any cross sections based on the observed
densities of isolated field dwarfs in the Local Group.  The same conclusion holds for
Milky Way satellites if one considers Figure 8 of \citet{Vogelsberger12}: SIDM with 
$\sigma/m = 10 ~\cmg$ (on the velocity scale of dwarf galaxies) can match the spread 
of local densities seen for the classical satellites of the Milky Way.  A larger simulation suite that includes a range of halo masses and more precise observational mass measurements may eventually allow such a constraint to be derived. 

Overall, we emphasize that the puzzling ``common mass" relation inferred from
collisionless CDM simulations, where luminous galaxies are no more dense, and 
potentially less dense, than galaxies 1000 times dimmer \citep[as is seen in the 
Local Group, e.g.][]{Strigari08,MBK12,GarrisonKimmelTBTF}, is a natural consequence 
of SIDM, where halo core densities generally increase with decreasing halo mass.


\section{Conclusions}
\label{sec:conclusions}

In this work, we  have used very high resolution cosmological simulations of 
$V_{\rm max} = 40 ~\kms$ halos to investigate the effect of SIDM on their 
density structure.  By simulating a range of SIDM cross sections and comparing our
simulations to the observed masses of local field dwarfs we have reached the 
following conclusions:

\begin{itemize}
\item SIDM models with $\sigma/\text{m} \simeq 0.5 - 50 ~\cmg$ on the velocity 
scale of dwarf galaxies ($v_{\rm rms} \simeq 40~\kms$) alleviate the TBTF problem 
and produce constant-density core profiles comparable in size to the half-light 
radii of Local Field dwarfs.  It is possible that cross sections even larger than 
$50 ~\cmg$ will alleviate the problem without producing catastrophic core collapse.

\item The largest ($\sim1$~kpc), lowest density ($\sim 0.03~\msun\,{\rm pc}^{-3}$) 
cores occur for models with $\sigma/\text{m} \simeq 5-10 ~\cmg$.  Our single run 
with $\sigma/\text{m} = 50 ~\cmg$ produces a slightly denser core owing to a mild 
degree of core collapse, yet it retains a constant-density profile at small radii 
and remains significantly less dense than the CDM case.

\item SIDM halo core densities increase inversely with halo circular velocity, roughly 
as $\rho_{\rm core} \propto 1/V_{\rm max}$ (Figure 5).  This fact may help explain the
unusual trend for less luminous dwarf galaxies to be denser or as dense as galaxies 
1000 times brighter \citep{Strigari08}.   However, this behavior also makes it difficult 
to rule out SIDM models by requiring halos to be at least as dense as the densest known
dwarfs (e.g. And~XVIII or Draco) since the host $V_{\rm max}$ is not well constrained.  

\end{itemize}

A much larger number of simulations, coupled with more precise measurements of dwarf 
galaxy densities, are required to place tight constraints on $\sigma/\text{m}$ at 
these velocity scales.  Further research, which is currently underway 
(\citealp{Vogelsberger14}; Elbert et al., in preparation; Robles et al., in preparation),
is also needed to investigate the effects of baryons on density profiles in an SIDM universe.  Overall, however, our results suggest that a wide range of SIDM cross sections remain viable on the velocity scales of dwarf galaxies, and that the range of cross sections that can alleviate TBTF spans at least two orders of magnitude.

\vskip1cm

\noindent {\bf{Acknowledgments}} \\
The authors thank Manoj Kaplinghat for helpful discussions.  
Support for this work was provided by NASA through \textit{Hubble Space Telescope} grants HST-GO-12966.003-A and HST-GO-13343.009-A.  
This work used the Extreme Science and Engineering Discovery Environment 
(XSEDE), which is supported by National Science Foundation grant number 
ACI-1053575.  We also acknowledge the computational support of the
\textit{Greenplanet} cluster at UCI, upon which much of the secondary 
analysis was performed.

\nocite{xsede}
\bibliography{Dwarf_SIDM}

\begin{thebibliography}{93}
\expandafter\ifx\csname natexlab\endcsname\relax\def\natexlab#1{#1}\fi

\bibitem[{{Amorisco} {et~al}\mbox{.}(2014){Amorisco}, {Zavala}, \& {de
  Boer}}]{Amorisco2014}
{Amorisco} N.~C., {Zavala} J., {de Boer} T.~J.~L., 2014, \apjl, 782, L39

\bibitem[{{Anderhalden} {et~al}\mbox{.}(2013){Anderhalden}, {Schneider},
  {Macci{\`o}}, {Diemand}, \& {Bertone}}]{Anderhalden13}
{Anderhalden} D., {Schneider} A., {Macci{\`o}} A.~V., {Diemand} J., {Bertone}
  G., 2013, \jcap, 3, 14

\bibitem[{{Arraki} {et~al}\mbox{.}(2014){Arraki}, {Klypin}, {More}, \&
  {Trujillo-Gomez}}]{Arraki14}
{Arraki} K.~S., {Klypin} A., {More} S., {Trujillo-Gomez} S., 2014, \mnras, 438,
  1466

\bibitem[{{Balberg} {et~al}\mbox{.}(2002){Balberg}, {Shapiro}, \&
  {Inagaki}}]{Balberg02}
{Balberg} S., {Shapiro} S.~L., {Inagaki} S., 2002, \apj, 568, 475

\bibitem[{{Behroozi} {et~al}\mbox{.}(2013){Behroozi}, {Wechsler}, \&
  {Wu}}]{Behroozi13}
{Behroozi} P.~S., {Wechsler} R.~H., {Wu} H.-Y., 2013, \apj, 762, 109

\bibitem[{{Boddy} {et~al}\mbox{.}(2014{\natexlab{a}}){Boddy}, {Feng},
  {Kaplinghat}, {Shadmi}, \& {Tait}}]{Boddy14a}
{Boddy} K.~K., {Feng} J.~L., {Kaplinghat} M., {Shadmi} Y., {Tait} T.~M.~P.,
  2014{\natexlab{a}}, {arXiv: 1408.6532 [hep-ph]}

\bibitem[{{Boddy} {et~al}\mbox{.}(2014{\natexlab{b}}){Boddy}, {Feng},
  {Kaplinghat}, \& {Tait}}]{Boddy14b}
{Boddy} K.~K., {Feng} J.~L., {Kaplinghat} M., {Tait} T.~M.~P.,
  2014{\natexlab{b}}, \prd, 89, 115017

\bibitem[{{Boylan-Kolchin} {et~al}\mbox{.}(2011){Boylan-Kolchin}, {Bullock}, \&
  {Kaplinghat}}]{MBK11}
{Boylan-Kolchin} M., {Bullock} J.~S., {Kaplinghat} M., 2011, \mnras, 415, L40

\bibitem[{{Boylan-Kolchin} {et~al}\mbox{.}(2012){Boylan-Kolchin}, {Bullock}, \&
  {Kaplinghat}}]{MBK12}
{Boylan-Kolchin} M., {Bullock} J.~S., {Kaplinghat} M., 2012, \mnras, 422, 1203

\bibitem[{{Brooks} \& {Zolotov}(2014)}]{Brooks2014}
{Brooks} A.~M., {Zolotov} A., 2014, \apj, 786, 87

\bibitem[{{Bryan} \& {Norman}(1998)}]{Bryan98}
{Bryan} G.~L., {Norman} M.~L., 1998, \apj, 495, 80

\bibitem[{{Buckley} {et~al}\mbox{.}(2014){Buckley}, {Zavala}, {Cyr-Racine},
  {Sigurdson}, \& {Vogelsberger}}]{Buckley2014}
{Buckley} M.~R., {Zavala} J., {Cyr-Racine} F.-Y., {Sigurdson} K.,
  {Vogelsberger} M., 2014, \prd, 90, 043524

\bibitem[{{Burkert}(1995)}]{Burkert95}
{Burkert} A., 1995, \apjl, 447, L25

\bibitem[{{Burkert}(2000)}]{Burkert2000}
{Burkert} A., 2000, \apjl, 534, L143

\bibitem[{{Cautun} {et~al}\mbox{.}(2014){Cautun}, {Frenk}, {van de Weygaert},
  {Hellwing}, \& {Jones}}]{Cautun2014}
{Cautun} M., {Frenk} C.~S., {van de Weygaert} R., {Hellwing} W.~A., {Jones}
  B.~J.~T., 2014, \mnras, 445, 2049

\bibitem[{{Col{\'{\i}}n} {et~al}\mbox{.}(2002){Col{\'{\i}}n}, {Avila-Reese},
  {Valenzuela}, \& {Firmani}}]{Colin02}
{Col{\'{\i}}n} P., {Avila-Reese} V., {Valenzuela} O., {Firmani} C., 2002, \apj,
  581, 777

\bibitem[{{Collins} {et~al}\mbox{.}(2013){Collins}, {Chapman}, {Rich}, {Ibata},
  {Martin}, {Irwin}, {Bate}, {Lewis}, {Pe{\~n}arrubia}, {Arimoto}, {Casey},
  {Ferguson}, {Koch}, {McConnachie}, \& {Tanvir}}]{Collins13}
{Collins} M.~L.~M. {et~al.}, 2013, \apj, 768, 172

\bibitem[{{Dav{\'e}} {et~al}\mbox{.}(2001){Dav{\'e}}, {Spergel}, {Steinhardt},
  \& {Wandelt}}]{Dave2001}
{Dav{\'e}} R., {Spergel} D.~N., {Steinhardt} P.~J., {Wandelt} B.~D., 2001,
  \apj, 547, 574

\bibitem[{{Del Popolo}(2012)}]{DelPopolo2012}
{Del Popolo} A., 2012, \mnras, 419, 971

\bibitem[{{Del Popolo} {et~al}\mbox{.}(2014){Del Popolo}, {Lima}, {Fabris}, \&
  {Rodrigues}}]{DelPopolo2014}
{Del Popolo} A., {Lima} J.~A.~S., {Fabris} J.~C., {Rodrigues} D.~C., 2014,
  \jcap, 4, 21

\bibitem[{{Di Cintio} {et~al}\mbox{.}(2014){Di Cintio}, {Brook}, {Dutton},
  {Macci{\`o}}, {Stinson}, \& {Knebe}}]{DiCintio14}
{Di Cintio} A., {Brook} C.~B., {Dutton} A.~A., {Macci{\`o}} A.~V., {Stinson}
  G.~S., {Knebe} A., 2014, \mnras, 441, 2986

\bibitem[{{Diemand} {et~al}\mbox{.}(2008){Diemand}, {Kuhlen}, {Madau}, {Zemp},
  {Moore}, {Potter}, \& {Stadel}}]{Diemand2008}
{Diemand} J., {Kuhlen} M., {Madau} P., {Zemp} M., {Moore} B., {Potter} D.,
  {Stadel} J., 2008, \nat, 454, 735

\bibitem[{{Dubinski} \& {Carlberg}(1991)}]{Dubinski1991}
{Dubinski} J., {Carlberg} R.~G., 1991, \apj, 378, 496

\bibitem[{{Epinat} {et~al}\mbox{.}(2008){Epinat}, {Amram}, {Marcelin},
  {Balkowski}, {Daigle}, {Hernandez}, {Chemin}, {Carignan}, {Gach}, \&
  {Balard}}]{Epinat08}
{Epinat} B. {et~al.}, 2008, \mnras, 388, 500

\bibitem[{{Ferrero} {et~al}\mbox{.}(2012){Ferrero}, {Abadi}, {Navarro},
  {Sales}, \& {Gurovich}}]{Ferrero2012}
{Ferrero} I., {Abadi} M.~G., {Navarro} J.~F., {Sales} L.~V., {Gurovich} S.,
  2012, \mnras, 425, 2817

\bibitem[{{Flores} \& {Primack}(1994)}]{Flores1994}
{Flores} R.~A., {Primack} J.~R., 1994, \apjl, 427, L1

\bibitem[{{Fraternali} {et~al}\mbox{.}(2009){Fraternali}, {Tolstoy}, {Irwin},
  \& {Cole}}]{Fraternali09}
{Fraternali} F., {Tolstoy} E., {Irwin} M.~J., {Cole} A.~A., 2009, \aap, 499,
  121

\bibitem[{{Garrison-Kimmel} {et~al}\mbox{.}(2014b){Garrison-Kimmel},
  {Boylan-Kolchin}, {Bullock}, \& {Kirby}}]{GarrisonKimmelTBTF}
{Garrison-Kimmel} S., {Boylan-Kolchin} M., {Bullock} J.~S., {Kirby} E.~N.,
  2014b, \mnras, 444, 222

\bibitem[{{Garrison-Kimmel} {et~al}\mbox{.}(2014a){Garrison-Kimmel},
  {Boylan-Kolchin}, {Bullock}, \& {Lee}}]{ELVIS}
{Garrison-Kimmel} S., {Boylan-Kolchin} M., {Bullock} J.~S., {Lee} K., 2014a,
  \mnras, 438, 2578

\bibitem[{{Garrison-Kimmel} {et~al}\mbox{.}(2014c){Garrison-Kimmel},
  {Horiuchi}, {Abazajian}, {Bullock}, \& {Kaplinghat}}]{GKBicep}
{Garrison-Kimmel} S., {Horiuchi} S., {Abazajian} K.~N., {Bullock} J.~S.,
  {Kaplinghat} M., 2014c, \mnras, 444, 961

\bibitem[{{Garrison-Kimmel} {et~al}\mbox{.}(2013){Garrison-Kimmel}, {Rocha},
  {Boylan-Kolchin}, {Bullock}, \& {Lally}}]{SGK13}
{Garrison-Kimmel} S., {Rocha} M., {Boylan-Kolchin} M., {Bullock} J.~S., {Lally}
  J., 2013, \mnras, 433, 3539

\bibitem[{{Gnedin} \& {Ostriker}(2001)}]{Gnedin01}
{Gnedin} O.~Y., {Ostriker} J.~P., 2001, \apj, 561, 61

\bibitem[{{Governato} {et~al}\mbox{.}(2012){Governato}, {Zolotov}, {Pontzen},
  {Christensen}, {Oh}, {Brooks}, {Quinn}, {Shen}, \& {Wadsley}}]{Governato12}
{Governato} F. {et~al.}, 2012, \mnras, 422, 1231

\bibitem[{{Griest}(1988)}]{Griest88}
{Griest} K., 1988, \prd, 38, 2357

\bibitem[{{Gritschneder} \& {Lin}(2013)}]{Gritschneder2013}
{Gritschneder} M., {Lin} D.~N.~C., 2013, \apj, 765, 38

\bibitem[{{Hahn} \& {Abel}(2011)}]{Hahn11}
{Hahn} O., {Abel} T., 2011, \mnras, 415, 2101

\bibitem[{{Hinshaw} {et~al}\mbox{.}(2013){Hinshaw}, {Larson}, {Komatsu},
  {Spergel}, {Bennett}, {Dunkley}, {Nolta}, {Halpern}, {Hill}, {Odegard},
  {Page}, {Smith}, {Weiland}, {Gold}, {Jarosik}, {Kogut}, {Limon}, {Meyer},
  {Tucker}, {Wollack}, \& {Wright}}]{WMAP9}
{Hinshaw} G. {et~al.}, 2013, \apjs, 208, 19

\bibitem[{{Hoffman} {et~al}\mbox{.}(1996){Hoffman}, {Salpeter}, {Farhat},
  {Roos}, {Williams}, \& {Helou}}]{Hoffman96}
{Hoffman} G.~L., {Salpeter} E.~E., {Farhat} B., {Roos} T., {Williams} H.,
  {Helou} G., 1996, \apjs, 105, 269

\bibitem[{{Horiuchi} {et~al}\mbox{.}(2014){Horiuchi}, {Humphrey}, {O{\~n}orbe},
  {Abazajian}, {Kaplinghat}, \& {Garrison-Kimmel}}]{Horiuchi14}
{Horiuchi} S., {Humphrey} P.~J., {O{\~n}orbe} J., {Abazajian} K.~N.,
  {Kaplinghat} M., {Garrison-Kimmel} S., 2014, \prd, 89, 025017

\bibitem[{{Jungman} {et~al}\mbox{.}(1996){Jungman}, {Kamionkowski}, \&
  {Griest}}]{Jungman96}
{Jungman} G., {Kamionkowski} M., {Griest} K., 1996, \physrep, 267, 195

\bibitem[{{Kaplinghat} {et~al}\mbox{.}(2014){Kaplinghat}, {Tulin}, \&
  {Yu}}]{Kaplinghat14}
{Kaplinghat} M., {Tulin} S., {Yu} H.-B., 2014, \prd, 89, 035009

\bibitem[{{Katz} \& {White}(1993)}]{Katz1993}
{Katz} N., {White} S.~D.~M., 1993, \apj, 412, 455

\bibitem[{{Kirby} {et~al}\mbox{.}(2014){Kirby}, {Bullock}, {Boylan-Kolchin},
  {Kaplinghat}, \& {Cohen}}]{Kirby14}
{Kirby} E.~N., {Bullock} J.~S., {Boylan-Kolchin} M., {Kaplinghat} M., {Cohen}
  J.~G., 2014, \mnras, 439, 1015

\bibitem[{{Klypin} {et~al}\mbox{.}(2014){Klypin}, {Karachentsev}, {Makarov}, \&
  {Nasonova}}]{Klypin2014}
{Klypin} A., {Karachentsev} I., {Makarov} D., {Nasonova} O., 2014, {arXiv:
  1405.4523 [astro-ph]}

\bibitem[{{Koch} {et~al}\mbox{.}(2007){Koch}, {Kleyna}, {Wilkinson}, {Grebel},
  {Gilmore}, {Evans}, {Wyse}, \& {Harbeck}}]{Koch2007}
{Koch} A., {Kleyna} J.~T., {Wilkinson} M.~I., {Grebel} E.~K., {Gilmore} G.~F.,
  {Evans} N.~W., {Wyse} R.~F.~G., {Harbeck} D.~R., 2007, \aj, 134, 566

\bibitem[{{Kochanek} \& {White}(2000)}]{Kochanek2000}
{Kochanek} C.~S., {White} M., 2000, \apj, 543, 514

\bibitem[{{Koda} \& {Shapiro}(2011)}]{Koda11}
{Koda} J., {Shapiro} P.~R., 2011, \mnras, 415, 1125

\bibitem[{{Komatsu} {et~al}\mbox{.}(2011){Komatsu}, {Smith}, {Dunkley},
  {Bennett}, {Gold}, {Hinshaw}, {Jarosik}, {Larson}, {Nolta}, {Page},
  {Spergel}, {Halpern}, {Hill}, {Kogut}, {Limon}, {Meyer}, {Odegard}, {Tucker},
  {Weiland}, {Wollack}, \& {Wright}}]{Komatsu11}
{Komatsu} E. {et~al.}, 2011, \apjs, 192, 18

\bibitem[{{Kuzio de Naray} {et~al}\mbox{.}(2008){Kuzio de Naray}, {McGaugh}, \&
  {de Blok}}]{KuziodeNaray08}
{Kuzio de Naray} R., {McGaugh} S.~S., {de Blok} W.~J.~G., 2008, \apj, 676, 920

\bibitem[{{Leaman} {et~al}\mbox{.}(2012){Leaman}, {Venn}, {Brooks},
  {Battaglia}, {Cole}, {Ibata}, {Irwin}, {McConnachie}, {Mendel}, \&
  {Tolstoy}}]{Leaman12}
{Leaman} R. {et~al.}, 2012, \apj, 750, 33

\bibitem[{{Loeb} \& {Weiner}(2011)}]{Loeb11}
{Loeb} A., {Weiner} N., 2011, Physical Review Letters, 106, 171302

\bibitem[{{Lovell} {et~al}\mbox{.}(2014){Lovell}, {Frenk}, {Eke}, {Jenkins},
  {Gao}, \& {Theuns}}]{Lovell14}
{Lovell} M.~R., {Frenk} C.~S., {Eke} V.~R., {Jenkins} A., {Gao} L., {Theuns}
  T., 2014, \mnras, 439, 300

\bibitem[{{Mateo} {et~al}\mbox{.}(2008){Mateo}, {Olszewski}, \&
  {Walker}}]{Mateo2008}
{Mateo} M., {Olszewski} E.~W., {Walker} M.~G., 2008, \apj, 675, 201

\bibitem[{{Mu{\~n}oz} {et~al}\mbox{.}(2005){Mu{\~n}oz}, {Frinchaboy},
  {Majewski}, {Kuhn}, {Chou}, {Palma}, {Sohn}, {Patterson}, \&
  {Siegel}}]{Munoz2005}
{Mu{\~n}oz} R.~R. {et~al.}, 2005, \apjl, 631, L137

\bibitem[{{Navarro} {et~al}\mbox{.}(1996){Navarro}, {Eke}, \&
  {Frenk}}]{Navarro96}
{Navarro} J.~F., {Eke} V.~R., {Frenk} C.~S., 1996, \mnras, 283, L72

\bibitem[{{Navarro} {et~al}\mbox{.}(1997){Navarro}, {Frenk}, \& {White}}]{NFW}
{Navarro} J.~F., {Frenk} C.~S., {White} S.~D.~M., 1997, \apj, 490, 493

\bibitem[{{O{\~n}orbe} {et~al}\mbox{.}(2014){O{\~n}orbe}, {Garrison-Kimmel},
  {Maller}, {Bullock}, {Rocha}, \& {Hahn}}]{Onorbe14}
{O{\~n}orbe} J., {Garrison-Kimmel} S., {Maller} A.~H., {Bullock} J.~S., {Rocha}
  M., {Hahn} O., 2014, \mnras, 437, 1894

\bibitem[{{Oh} {et~al}\mbox{.}(2008){Oh}, {de Blok}, {Walter}, {Brinks}, \&
  {Kennicutt}}]{Oh08}
{Oh} S.-H., {de Blok} W.~J.~G., {Walter} F., {Brinks} E., {Kennicutt}, Jr.
  R.~C., 2008, \aj, 136, 2761

\bibitem[{{Papastergis} {et~al}\mbox{.}(2014){Papastergis}, {Giovanelli},
  {Haynes}, \& {Shankar}}]{Papastergis14}
{Papastergis} E., {Giovanelli} R., {Haynes} M.~P., {Shankar} F., 2014, {arXiv:
  1407.4665 [astro-ph]}

\bibitem[{{Pe{\~n}arrubia} {et~al}\mbox{.}(2012){Pe{\~n}arrubia}, {Pontzen},
  {Walker}, \& {Koposov}}]{Penarrubia12}
{Pe{\~n}arrubia} J., {Pontzen} A., {Walker} M.~G., {Koposov} S.~E., 2012,
  \apjl, 759, L42

\bibitem[{{Peter} {et~al}\mbox{.}(2013){Peter}, {Rocha}, {Bullock}, \&
  {Kaplinghat}}]{Peter13}
{Peter} A.~H.~G., {Rocha} M., {Bullock} J.~S., {Kaplinghat} M., 2013, \mnras,
  430, 105

\bibitem[{{Planck Collaboration} {et~al}\mbox{.}(2014){Planck Collaboration},
  {Ade}, {Aghanim}, {Armitage-Caplan}, {Arnaud}, {Ashdown}, {Atrio-Barandela},
  {Aumont}, {Baccigalupi}, {Banday}, \& et~al.}]{PlanckCos}
{Planck Collaboration} {et~al.}, 2014, \aap, 571, A16

\bibitem[{{Polisensky} \& {Ricotti}(2014)}]{PolisenskynRicotti}
{Polisensky} E., {Ricotti} M., 2014, \mnras, 437, 2922

\bibitem[{{Pontzen} \& {Governato}(2012)}]{Pontzen12}
{Pontzen} A., {Governato} F., 2012, \mnras, 421, 3464

\bibitem[{{Power} {et~al}\mbox{.}(2003){Power}, {Navarro}, {Jenkins}, {Frenk},
  {White}, {Springel}, {Stadel}, \& {Quinn}}]{Power03}
{Power} C., {Navarro} J.~F., {Jenkins} A., {Frenk} C.~S., {White} S.~D.~M.,
  {Springel} V., {Stadel} J., {Quinn} T., 2003, \mnras, 338, 14

\bibitem[{{Purcell} \& {Zentner}(2012)}]{Purcell12}
{Purcell} C.~W., {Zentner} A.~R., 2012, \jcap, 12, 7

\bibitem[{{Randall} {et~al}\mbox{.}(2008){Randall}, {Markevitch}, {Clowe},
  {Gonzalez}, \& {Brada{\v c}}}]{Randall08}
{Randall} S.~W., {Markevitch} M., {Clowe} D., {Gonzalez} A.~H., {Brada{\v c}}
  M., 2008, \apj, 679, 1173

\bibitem[{{Reid} {et~al}\mbox{.}(2010){Reid}, {Percival}, {Eisenstein},
  {Verde}, {Spergel}, {Skibba}, {Bahcall}, {Budavari}, {Frieman}, {Fukugita},
  {Gott}, {Gunn}, {Ivezi{\'c}}, {Knapp}, {Kron}, {Lupton}, {McKay}, {Meiksin},
  {Nichol}, {Pope}, {Schlegel}, {Schneider}, {Stoughton}, {Strauss}, {Szalay},
  {Tegmark}, {Vogeley}, {Weinberg}, {York}, \& {Zehavi}}]{Reid10}
{Reid} B.~A. {et~al.}, 2010, \mnras, 404, 60

\bibitem[{{Rocha} {et~al}\mbox{.}(2013){Rocha}, {Peter}, {Bullock},
  {Kaplinghat}, {Garrison-Kimmel}, {O{\~n}orbe}, \& {Moustakas}}]{Rocha13}
{Rocha} M., {Peter} A.~H.~G., {Bullock} J.~S., {Kaplinghat} M.,
  {Garrison-Kimmel} S., {O{\~n}orbe} J., {Moustakas} L.~A., 2013, \mnras, 430,
  81

\bibitem[{{Rodr{\'{\i}}guez-Puebla}
  {et~al}\mbox{.}(2013){Rodr{\'{\i}}guez-Puebla}, {Avila-Reese}, \&
  {Drory}}]{Rodriguez-Puebla13}
{Rodr{\'{\i}}guez-Puebla} A., {Avila-Reese} V., {Drory} N., 2013, \apj, 773,
  172

\bibitem[{{Simon} \& {Geha}(2007)}]{Simon2007}
{Simon} J.~D., {Geha} M., 2007, \apj, 670, 313

\bibitem[{{Spergel} \& {Steinhardt}(2000)}]{Spergel00}
{Spergel} D.~N., {Steinhardt} P.~J., 2000, Physical Review Letters, 84, 3760

\bibitem[{{Springel}(2005)}]{Springel05}
{Springel} V., 2005, \mnras, 364, 1105

\bibitem[{{Springel} {et~al}\mbox{.}(2008){Springel}, {Wang}, {Vogelsberger},
  {Ludlow}, {Jenkins}, {Helmi}, {Navarro}, {Frenk}, \& {White}}]{Aquarius}
{Springel} V. {et~al.}, 2008, \mnras, 391, 1685

\bibitem[{{Steigman} \& {Turner}(1985)}]{SteigmanTurner85}
{Steigman} G., {Turner} M.~S., 1985, Nuclear Physics B, 253, 375

\bibitem[{{Strigari} {et~al}\mbox{.}(2008){Strigari}, {Bullock}, {Kaplinghat},
  {Simon}, {Geha}, {Willman}, \& {Walker}}]{Strigari08}
{Strigari} L.~E., {Bullock} J.~S., {Kaplinghat} M., {Simon} J.~D., {Geha} M.,
  {Willman} B., {Walker} M.~G., 2008, \nat, 454, 1096

\bibitem[{{Strigari} {et~al}\mbox{.}(2014){Strigari}, {Frenk}, \&
  {White}}]{Strigari14}
{Strigari} L.~E., {Frenk} C.~S., {White} S.~D.~M., 2014, {arXiv: 1406.6079
  [astro-ph]}

\bibitem[{{Tollerud} {et~al}\mbox{.}(2014){Tollerud}, {Boylan-Kolchin}, \&
  {Bullock}}]{Tollerud14}
{Tollerud} E.~J., {Boylan-Kolchin} M., {Bullock} J.~S., 2014, \mnras, 440, 3511

\bibitem[{Towns {et~al}\mbox{.}(2014)Towns, Cockerill, Dahan, Foster, Gaither,
  Grimshaw, Hazlewood, Lathrop, Lifka, Peterson, Wilkens-Diehr, Roskies, \&
  Scott}]{xsede}
Towns J. {et~al.}, 2014, Computing in Science and Engineering, 99, 1

\bibitem[{{Tulin} {et~al}\mbox{.}(2013{\natexlab{a}}){Tulin}, {Yu}, \&
  {Zurek}}]{TulinYuZurek}
{Tulin} S., {Yu} H.-B., {Zurek} K.~M., 2013{\natexlab{a}}, \prd, 87, 115007

\bibitem[{{Tulin} {et~al}\mbox{.}(2013{\natexlab{b}}){Tulin}, {Yu}, \&
  {Zurek}}]{Tulin13}
{Tulin} S., {Yu} H.-B., {Zurek} K.~M., 2013{\natexlab{b}}, Physical Review
  Letters, 110, 111301

\bibitem[{{Vogelsberger} \& {Zavala}(2013)}]{Vogelsberger13}
{Vogelsberger} M., {Zavala} J., 2013, \mnras, 430, 1722

\bibitem[{{Vogelsberger} {et~al}\mbox{.}(2012){Vogelsberger}, {Zavala}, \&
  {Loeb}}]{Vogelsberger12}
{Vogelsberger} M., {Zavala} J., {Loeb} A., 2012, \mnras, 423, 3740

\bibitem[{{Vogelsberger} {et~al}\mbox{.}(2014){Vogelsberger}, {Zavala},
  {Simpson}, \& {Jenkins}}]{Vogelsberger14}
{Vogelsberger} M., {Zavala} J., {Simpson} C., {Jenkins} A., 2014, \mnras, 444,
  3684

\bibitem[{{Walker} {et~al}\mbox{.}(2009){Walker}, {Mateo}, \&
  {Olszewski}}]{Walker2009}
{Walker} M.~G., {Mateo} M., {Olszewski} E.~W., 2009, \aj, 137, 3100

\bibitem[{{Walker} \& {Pe{\~n}arrubia}(2011)}]{Walker11}
{Walker} M.~G., {Pe{\~n}arrubia} J., 2011, \apj, 742, 20

\bibitem[{{Wang} {et~al}\mbox{.}(2012){Wang}, {Frenk}, {Navarro}, {Gao}, \&
  {Sawala}}]{Wang2012}
{Wang} J., {Frenk} C.~S., {Navarro} J.~F., {Gao} L., {Sawala} T., 2012, \mnras,
  424, 2715

\bibitem[{{Weiner} {et~al}\mbox{.}(2006){Weiner}, {Willmer}, {Faber},
  {Melbourne}, {Kassin}, {Phillips}, {Harker}, {Metevier}, {Vogt}, \&
  {Koo}}]{Weiner2006}
{Weiner} B.~J. {et~al.}, 2006, \apj, 653, 1027

\bibitem[{{Wolf} {et~al}\mbox{.}(2010){Wolf}, {Martinez}, {Bullock},
  {Kaplinghat}, {Geha}, {Mu{\~n}oz}, {Simon}, \& {Avedo}}]{Wolf10}
{Wolf} J., {Martinez} G.~D., {Bullock} J.~S., {Kaplinghat} M., {Geha} M.,
  {Mu{\~n}oz} R.~R., {Simon} J.~D., {Avedo} F.~F., 2010, \mnras, 406, 1220

\bibitem[{{Yoshida} {et~al}\mbox{.}(2000){Yoshida}, {Springel}, {White}, \&
  {Tormen}}]{Yoshida00b}
{Yoshida} N., {Springel} V., {White} S.~D.~M., {Tormen} G., 2000, \apjl, 544,
  L87

\bibitem[{{Zavala} {et~al}\mbox{.}(2013){Zavala}, {Vogelsberger}, \&
  {Walker}}]{Zavala13}
{Zavala} J., {Vogelsberger} M., {Walker} M.~G., 2013, \mnras, 431, L20

\bibitem[{{Zentner} \& {Bullock}(2002)}]{ZB02}
{Zentner} A.~R., {Bullock} J.~S., 2002, \prd, 66, 043003

\bibitem[{{Zolotov} {et~al}\mbox{.}(2012){Zolotov}, {Brooks}, {Willman},
  {Governato}, {Pontzen}, {Christensen}, {Dekel}, {Quinn}, {Shen}, \&
  {Wadsley}}]{Zolotov2012}
{Zolotov} A. {et~al.}, 2012, \apj, 761, 71

\end{thebibliography}

\appendix
\section{Velocity dispersion profiles}
\label{appendixA}
\begin{figure*}
\centering
\includegraphics[width =\columnwidth]{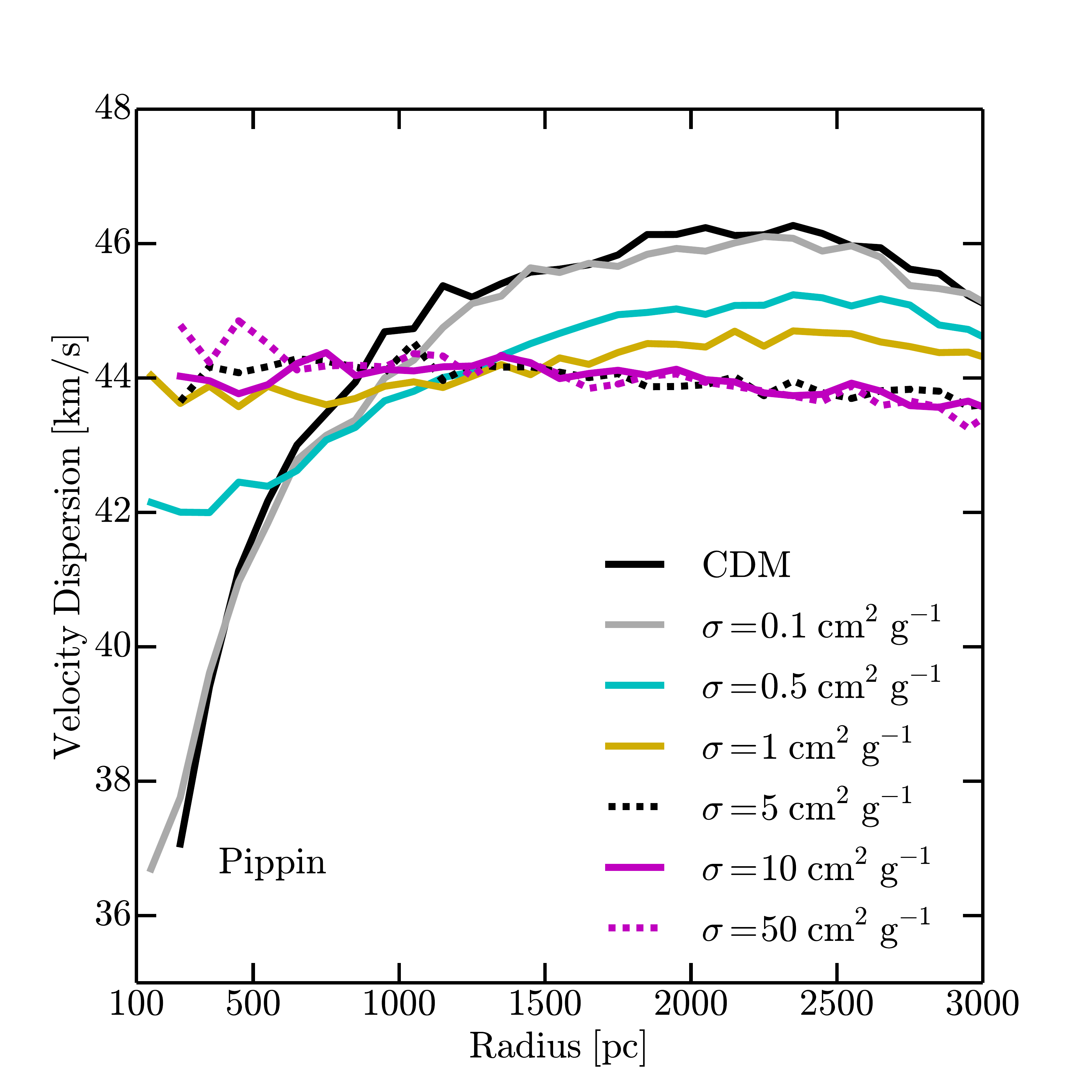} 
\includegraphics[width =\columnwidth]{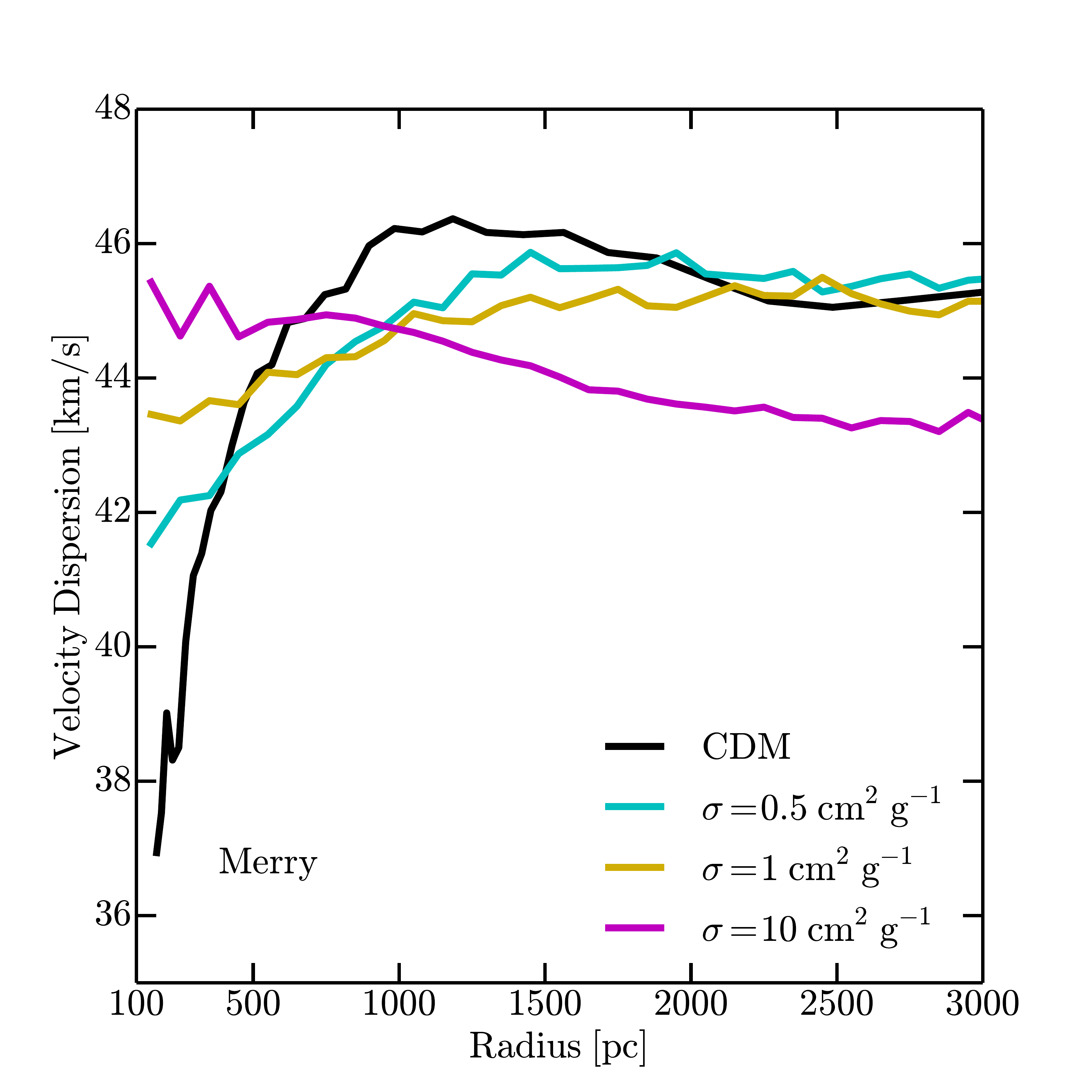}
\caption{Velocity dispersion profiles of Pippin (left) and Merry (right) in
collisionless CDM and with a variety of SIDM cross sections (see legend).  As
the cross section increases to $\sigmam = 10~\cmg$, the cores become increasingly
hotter and the profiles become more isothermal as kinetic energy is transported
to the center of the halo.  Very high cross sections $\sigmam \geq 10~\cmg$
lead to a negative radial gradient, resulting in energy transport from the
center of the halo to the outer parts, resulting in mild core collapse and
higher central densities, as in the $50~\cmg$ run plotted in 
Figure~\ref{fig:profiles}.}
\label{fig:disp}
\end{figure*}

The velocity dispersion profiles of SIDM halos provide insight into the origin 
and nature of their cored density structures.  Figure~\ref{fig:disp} plots the 
velocity dispersion profiles of our halos both in collisionless CDM and in SIDM, 
in direct analogy to the density profiles shown in Figure~\ref{fig:profiles}.  As 
the SIDM cross section is increased from $\sigma/m = 0.1 \rightarrow 5-10~\cmg$, 
the cores become steadily hotter and more isothermal as kinetic energy is 
transported from the outside in (resulting in increasingly lowered central 
densities relative to the cold cusp that forms in CDM).  However, the run with a 
cross section larger than $\sigma/m \sim 10~\cmg$ begins to display a negative 
radial gradient -- i.e., a core that is hotter than the outer regions.  
This is precisely the situation where core collapse behavior is expected
in SIDM:  heat is transferred out of the halo center, resulting in decreased 
pressure support and ultimately a density enhancement (and further heating).
This core collapse behavior is seen explicitly in the left panel of 
Figure~\ref{fig:profiles}, where the Pippin run with $\sigma/m = 50~\cmg$ is 
much denser than the $\sigma/m = 5, 10 ~\cmg$ cases.

\section{Comparisons to Classical Milky Way Satellites}
\label{appendixB}
\begin{figure}
\centering
\includegraphics[width = \columnwidth]{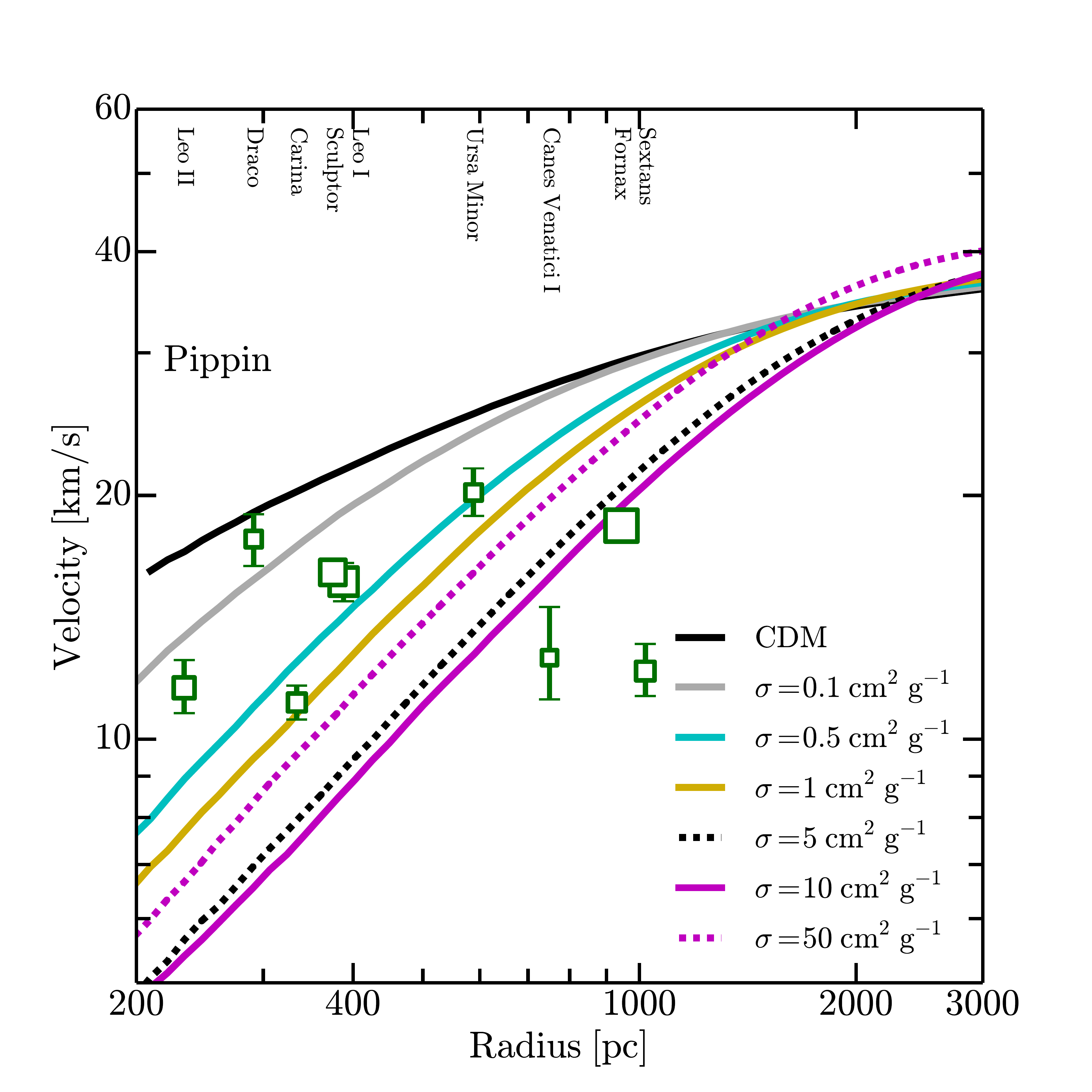}
\caption{Circular velocity profiles of Pippin as in 
Figure~\ref{fig:rotationcurves}, but with constraints on the circular 
velocities of the bright Milky Way satellites, from \citet{Wolf10}.  SIDM 
simulations with $\sigmam \geq 0.5~\cmg$ resolve TBTF, as in the field.
However, the simulated halos have not undergone the environmental processes
typical of subhalos and satellite galaxies; a quantitative comparison is
therefore impossible.}
\label{fig:mwvcirc}
\end{figure}

While Pippin and Merry are isolated dark matter halos, rather than subhalos, 
it is nonetheless interesting to compare their velocity profiles to the 
brightest Milky Way dwarf satellites.  Figure~\ref{fig:mwvcirc} plots the 
same circular velocity profiles of Pippin as in Figure~\ref{fig:rotationcurves}, but 
with the data points replaced by measurements of the circular velocities of 
the bright ($\mstar > 2\times10^5\msun$) Milky Way satellites used to define 
TBTF in \citet{MBK11,MBK12}.  The points are taken from \citet{Wolf10}, who
used data from \citet{Munoz2005}, \citet{Koch2007}, \citet{Simon2007},
\citet{Mateo2008}, and \citet{Walker2009}.  As in the field 
(Figure~\ref{fig:rotationcurves}), SIDM runs with $\sigmam \geq 0.5~\cmg$
alleviate TBTF significantly.  Because the simulated halos have not been tidally
stripped by a larger host halo, however, a quantitative comparison is impossible 
and we instead present these results for illustrative purposes only.


\label{lastpage}
\end{document}